%% file: main.tex
\documentclass[epj,referee]{svjour}

\usepackage{graphics}
\usepackage{enumerate}
\usepackage{amsmath}
\usepackage[percent]{overpic}

\usepackage[T1]{fontenc} 
\usepackage[utf8]{inputenc}

\usepackage{lineno}

\usepackage{xcolor}
\usepackage{colortbl}
\usepackage{multicol}
\usepackage[colorlinks=true,linkcolor=firebrick,citecolor=green, urlcolor=darkblue]{hyperref}
\definecolor{darkred}{rgb}{0.5,0,0}
\definecolor{darkblue}{rgb}{0,0,0.5}
\definecolor{firebrick}{rgb}{0.75,0.125,0.125}
\definecolor{darkgreen}{rgb}{0,0.5,0}

\usepackage[
    backend=biber,      
    style=numeric,      
    sorting=none,       
    giveninits=true,    
    url=true,           
]{biblatex}

\renewbibmacro{in:}{}

\DeclareFieldFormat{pages}{#1}

\usepackage{xspace}

\newcommand{\byc}{\kern-0.1em/\kern-0.1em c}
\newcommand{\TeV}{\ensuremath{\mbox{Te\kern-0.1em V}}\xspace}
\newcommand{\GeV}{\ensuremath{\mbox{Ge\kern-0.1em V}}\xspace}
\newcommand{\MeV}{\ensuremath{\mbox{Me\kern-0.1em V}}\xspace}
\newcommand{\GeVc}{\ensuremath{\mbox{Ge\kern-0.1em V}\byc}\xspace}
\newcommand{\AGeVc}{\ensuremath{A\,\mbox{Ge\kern-0.1em V}\byc}\xspace}
\newcommand{\MeVc}{\ensuremath{\mbox{Me\kern-0.1em V}\byc}\xspace}

\newcommand{\pt}{\ensuremath{p_{\textrm T}}\xspace}

\newcommand{\y}{\ensuremath{{y}}\xspace}
\newcommand{\p}{\ensuremath{{p}}\xspace}
\newcommand{\pp}{\mbox{\textit{p}+\textit{p}}\xspace}
\newcommand{\ee}{\mbox{\textit{e$^+$}+\textit{e$^-$}}\xspace}
\newcommand{\eD}{\mbox{\textit{e$^-$}+D}\xspace}
\newcommand{\snn}{\ensuremath{\sqrt{s_{\mathrm{NN}}}}\xspace}

\newcommand{\ks}{\ensuremath{K^0_\mathrm{S}}\xspace}

\newcommand{\pim}{\ensuremath{\pi^-}\xspace}
\newcommand{\pip}{\ensuremath{\pi^+}\xspace}
\newcommand{\km}{\ensuremath{\textit{K}^-}\xspace}
\newcommand{\kp}{\ensuremath{\textit{K}^+}\xspace}

\usepackage{comment}

\begin{document}
\onecolumn

\title{Isospin-symmetry violation - kaons and beyond}
\subtitle{ISO-BREAK 25: summary and outlook}
\author{
Marek Gazdzicki\inst{1} (editor),   
Francesco Giacosa\inst{1} (editor), 
Katarzyna Grebieszkow\inst{2} (editor), 
David Blaschke\inst{3},         
Marcus Bleicher\inst{4},        
Bastian Brandt\inst{5},         
Wojciech Bryliński\inst{},     
Tobiasz Czopowicz\inst{6},      
Jim Drachenberg\inst{7},        
Dipangkar Dutta\inst{8},        
Francesca Ercolessi\inst{9},   
Mark Gorenstein\inst{10},       
Linqin Huang\inst{11},          
Oleksii Ivanytskyi\inst{3},     
Nicol\`o Jacazio\inst{12},      
Joseph Kapusta\inst{13},        
Seweryn Kowalski\inst{14},      
Maciej Piotr Lewicki\inst{15},   
Manuel Lorenz\inst{16},         
Stanisław Mrówczyński\inst{6},  
Vitalii Ozvenchuk\inst{15},      
Oleksandra Panova\inst{1},      
Roman Płaneta\inst{17},         
Krzysztof Piasecki\inst{18},    
Milena Piotrowska\inst{1},      
Rob Pisarski\inst{19},          
Damian Pszczel\inst{6},         
Johann Rafelski\inst{20},       
Martin Rohrmoser\inst{1},       
Andrzej Rybicki\inst{15},        
Maciej Rybczyński\inst{1},      
Radoslaw Ryblewski\inst{15},     
Subhasis Samanta\inst{21},      
Mayank Singh\inst{22},          
Joanna Maria Stepaniak\inst{6}, 
Grzegorz Stefanek\inst{1},      
Herbert Str\"obele\inst{4},     
Tatjana {\v{S}}u{\v{s}}a\inst{23},  
Leonardo Tinti\inst{1},         
Ludwik Turko\inst{3},           
Oleksandr Vitiuk\inst{3},       
Klaus Werner\inst{24},          
Hanna Zbroszczyk\inst{2}        
}
\institute{
Jan Kochanowski University, Kielce, Poland \and
Warsaw University of Technology, Warsaw, Poland \and
University of Wroclaw, Poland \and
Goethe University, Frankfurt am Main, Germany \and
University of Bielefeld, Germany \and
National Centre for Nuclear Research, Warsaw, Poland \and
Abilene Christian University, USA \and
Mississippi State University, USA \and
Universit\`a e INFN, Bologna, Italy \and
Bogolyubov Institute for Theoretical Physics, Kyiv, Ukraine \and
Institute of Modern Physics, Chinese Academy of Sciences, China \and
Universit\`a del Piemonte Orientale, Italy \and
University of Minnesota, USA \and
University of Silesia, Poland \and
Institute of Nuclear Physics, Polish Academy of Sciences, Krakow, Poland \and 
GSI Helmholtzzentrum f\"ur Schwerionenforschung, Darmstadt, Germany \and
Jagiellonian University, Krakow, Poland \and
University of Warsaw, Poland \and
Brookhaven National Laboratory, USA \and
University of Arizona, Tucson, USA \and
Kalinga Institute of Industrial Technology, India \and
Vanderbilt University, USA \and
Ru{\dj}er Bo\v{s}kovi\'c Institute, Zagreb, Croatia \and
SUBATECH, Nantes University--IN2P3/CNRS--IMT Atlantique, Nantes, France
}

%
\date{Received: date / Revised version: date}
%
\abstract{
This report summarizes the presentations and discussions during the
ISO-BREAK 25 Workshop ``Isospin symmetry violation: kaons and beyond'', which was held at Jan Kochanowski University in Kielce on October 23--25, 2025. We address the current status of the isospin-symmetry breaking discovered by NA61/SHINE in nucleus--nucleus collisions at the CERN SPS, its confirmation by other experiments and studies in \ee and deep inelastic scattering. In addition, we discuss the theoretical status as well as we outline experimental and theoretical priorities towards understanding this currently unexplained phenomenon.
\PACS{
      {PACS-key}{discribing text of that key}   \and
      {PACS-key}{discribing text of that key}
     } 
} 

\authorrunning{M. Gazdzicki, F. Giacosa, K. Grebieszkow, et al.}
\maketitle
\newpage

\section{Introduction}
\label{intro}

\input{sections/Introduction}

\section{Experimental evidence for charge-symmetry violation}
\label{sec:data}

\input{sections/Data}

\section{Legacy model expectations}
\label{sec:legacy}
\input{sections/Th_legacy}

\section{Phenomenological descriptions versus understanding origins}
\label{sec:fitting}
\input{sections/Th_fit}

\section{Brainstorming possible explanations}
\label{sec:brain}
\input{sections/Brain}

\section{What next?}
\label{sec:next}
\input{sections/Next}

\bigskip
\textbf{Acknowledgments}

The organizers/editors acknowledge support by the Minister of Science (Poland) under the `Regional Excellence Initiative' program (project no.: RID/SP/0015/2024/01), Polish National Science Centre (NCN) grant MAESTRO 2018/30/A/ST2/00226, and 
the Polish Minister of Science and Higher Education (contract No. 2025/WK/05).
We thank other participants of the workshop, not included in the author list, for their discussions.

\bigskip
\textbf{International Advisory Committee:}
Marcus Bleicher,    
Bastian Brandt,    
Marek Gazdzicki,    
Francesco Giacosa,    
Mark Gorenstein,    
Katarzyna Grebieszkow,    
Joseph Kapusta,   
Rob Pisarski,    
Andrzej Rybicki,    
Tatjana {\v{S}}u{\v{s}}a,   
Hanna Zbroszczyk.   
\textbf{Organizers:} 
Tobiasz Czopowicz,
Marek Gazdzicki,
Francesco Giacosa,
Katarzyna Grebieszkow,
Milena Piotrowska,
Grzegorz Stefanek. 
\textbf{Local organizing support:} 
Krzysztof Kyzioł, 
Arthur Vereijken.

\bigskip
\textbf{Author contributions}

The main text of the paper was drafted by the editors (M.~Gazdzicki, F.~Giacosa, and K.~Grebieszkow) and then modified by the presenters of a given topic at the workshop.
The appendices were drafted by J.~Rafelski~(\ref{appR}) and S.~Mrówczyński~(\ref{appM}).
Figures~\ref{fig:NA61_pions}, \ref{fig:ALICE_RK}~(\textit{right}), and \ref{fig:models_compil} were prepared by K.~Grebieszkow, F.~Ercolessi and N.~Jacazio, and S.~Samanta, respectively.
The remaining figures are taken from the cited talks or publications.
The whole paper was reviewed and corrected by all authors.

\bigskip
\section*{Appendices}
\appendix

\section{Strong electromagnetic fields as a source of isospin breaking effects}
\label{appR}
\input{sections/appR}

\section{On the role of \texorpdfstring{$\pmb{u}$--$\pmb{d}$}{u--d} mass splitting in quark pair production}
\label{appM}

\input{sections/appM}

\printbibliography

\end{document}

%% file: sections/Introduction.tex
\begin{figure}[h!]
\centering

\resizebox{0.7\textwidth}{!}{
  \includegraphics{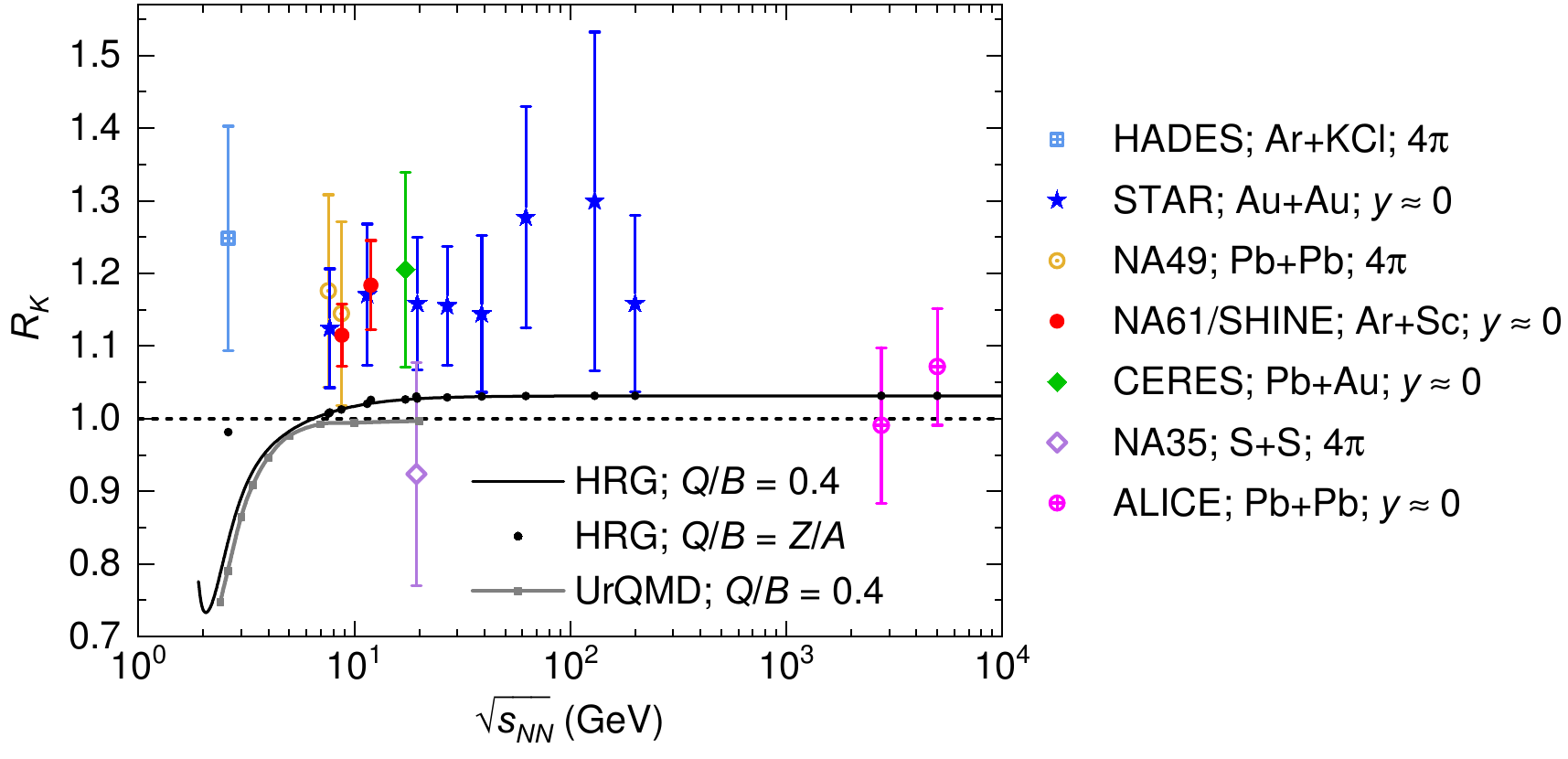}
}
\caption{Ratio of charged to neutral $K$ meson yields ($R_K=(\kp+\km)/2\ks$) in nucleus--nucleus collisions as a function of collision center-of-mass energy per nucleon pair. The black line shows the Hadron Resonance Gas (HRG) model predictions for $Q/B=0.4$, where $Q/B$ denotes the electric charge per baryon number of the whole system. The black dots indicate the HRG predictions for $Q/B$ values corresponding to the ones in the experiments. The gray squares show UrQMD predictions. Vertical bars denote total uncertainties. The plot is taken from Ref.~\cite{NA61SHINE:2023azp}, updated with the preliminary Ar+Sc 8.8~\GeV point~\cite{Balkova:2025dpf}, and the Pb+Pb 5.02~\TeV point from Ref.~\cite{Ercolessi_ISOBREAK25} with later changes. The numerical results used to obtain the $R_K$ values are taken from the
HADES~\mbox{\cite{HADES:2009lnd,Agakishiev:2010zw}},
STAR~\mbox{\cite{STAR:2017sal,STAR:2008med,STAR:2002hpr,STAR:2019bjj,STAR:2010yyv,STAR:2011fbd}}, 
NA49~\mbox{\cite{NA49:2007stj,NA49:2002pzu,Strabel:thesis}},
NA61/SHINE~\mbox{\cite{NA61SHINE:2023epu,NA61SHINE:2023azp,Balkova:2025dpf}},
CERES~\mbox{\cite{Kalisky:thesis,Radomski:2008zz,Radomski:thesis}}, 
NA35~\mbox{\cite{NA35:1992njd,NA35:1994veo}}, and
ALICE~\mbox{\cite{ALICE:2013mez,ALICE:2019hno,ALICE:2013cdo,ALICE:2025cqy}} experiments. Note that the CERES results~\mbox{\cite{Kalisky:thesis,Radomski:2008zz,Radomski:thesis}}, as well as the NA49 \ks results~\cite{Strabel:thesis} and NA61/SHINE \ks result at 8.8 \GeV~\cite{Balkova:2025dpf}, are preliminary.
}
\label{fig:NA61_RK}
\end{figure}

Recently, the NA61/SHINE Collaboration at the CERN SPS reported a significant excess of charged over neutral kaon production
in high-energy nucleus--nucleus collisions~\cite{NA61SHINE:2023azp}. 
The NA61/SHINE results, together with the compilation of the world data, are shown in Fig.~\ref{fig:NA61_RK}.
Exact isospin symmetry results in equal yields of charged and neutral kaons~\cite{Brylinski:2023nrb}. More narrowly, this prediction follows from charge symmetry, which is part of isospin symmetry. These symmetries are related, at the quark level, to $ud$-flavor symmetry.

At low energies, isospin symmetry is approximately obeyed by hadrons because the effects of the $u$--$d$ quark-mass difference and electromagnetic interactions are small on the hadronic scale, as reflected, for example, in the mass splittings within isospin multiplets ~\cite{ParticleDataGroup:2024cfk}. Accordingly, pion--pion and
pion--nucleon scatterings are often successfully treated in the
isospin-symmetric limit; see, e.g.,
Refs.~\cite{Pennington:2005be,Alarcon:2011zs}.
Explicit calculations that take into account isospin-symmetry breaking find small corrections in pion--nucleon \cite{Meissner:1997ii,Hoferichter:2009ez,Hoferichter:2009gn} and pion--pion scattering \cite{Colangelo:2025iuq}.
The same qualitative pattern is seen in the suppression of isospin-violating decays, such as
such as $\eta'(958)\to\pi\pi\pi$~\cite{Gross:1979ur,Serpukhov-Brussels-AnnecyLAPP:1984xib},
and $\rho(770)\to\pi\pi\pi$ and $\omega(782)\to\pi\pi$, associated
with the small $\rho$--$\omega$ mixing
\cite{OConnell:1995nse,Kovacs:2024cdb}. 
These examples confirm the expectation that low-energy isospin-breaking effects are small. They do not, however, provide a basis for quantitative predictions of the charged-to-neutral kaon yield ratio in high-energy collisions.

The NA61/SHINE Collaboration quantified the charge-symmetry violation (CSV) in the kaon sector by the ratio
\begin{equation}
    R_K = \frac{\kp + \km}{K^0+ \overline{K}^0} = \frac{\kp + \km}{2 \ks}
\end{equation}
reporting $R_K=1.184 \pm 0.061$~\cite{NA61SHINE:2023azp}. 
In the isospin-symmetric limit, $R_K=1$ follows from $\kp =K^0$ and $\km = \overline{K}^0$ imposed by charge-symmetry conservation, see Refs.~\cite{Brylinski:2023nrb,Giacosa:2024bup} for a detailed discussion. As theoretical models show (Fig.~\ref{fig:NA61_RK} and later on), the deviation from unity is expected to be small (at most a few percent). 
Soon after, an indication of unexplained CSV was reported 
in \ee collisions~\cite{BESIII:2025mbc} and
deep inelastic electron--deuteron scattering~\cite{Bhatt:2024prq}.

Thus, the magnitude of CSV was neither predicted nor can be explained by existing phenomenological approaches inspired by the Standard Model~\cite{Brylinski:2023nrb,BESIII:2025mbc,Bhatt:2024prq}. 
Three possible causes may contribute to it:
\begin{enumerate}[(i)]
    \item 
    The experimental data are incorrect (central values and/or their uncertainties).
    \item
    The models are incorrect.
    \item 
    Data and models are incorrect.
\end{enumerate}
The goal of the ISO-BREAK 25 Workshop, held in October 2025 in Kielce, was to identify the source of the disagreement between the data and the models regarding the magnitude of the charge-symmetry violation.
The answer is expected to have a significant impact on either the data analysis or the modeling of high-energy processes. \textit{In the last resort}, it may require the introduction of physics beyond the Standard Model.

This document is organized as follows. Section~\ref{sec:data} briefly reports experimental status. The predictions of models are summarized in Sec.~\ref{sec:legacy}. The observed asymmetry between charged and neutral kaon yields is fitted by several models by explicitly introducing charge-symmetry-violation parameters.
This is reviewed in Sec.~\ref{sec:fitting}. Section~\ref{sec:brain} lists ideas that may help understand the charge-symmetry violation at high energies. 
In particular, two appendices deepen the discussion of the role of 
strong electromagnetic fields in high-energy collisions and
different masses of $u$ and $d$ quarks in QCD $q\overline{q}$ creation. 
Section~\ref{sec:next} closes the report by discussing experimental and theoretical priorities.

%% file: sections/Data.tex
The charge-symmetry violation in strong interactions beyond known effects was first reported by NA61/SHINE in the production of kaons in Ar+Sc collisions at $\snn =$ 11.9~\GeV~\cite{NA61SHINE:2023azp}. This was an unexpected discovery.
Soon afterward, indications of CSV were also reported in \ee collisions at 3.0--3.7~\GeV and in deep inelastic scattering (DIS) of 10~\GeV electrons on deuterons (D).
In hindsight, this sequence is understandable: the prediction of charge symmetry for the charged-to-neutral kaon ratio in nucleus--nucleus collisions is only weakly model dependent and relatively straightforward to test experimentally~\cite{NA61SHINE:2023azp,Brylinski:2023nrb}.
In contrast, for processes involving electrons, separating charge-symmetry violation arising from electromagnetic interactions in the initial state from CSV in quark fragmentation into hadrons is significantly model dependent.
Here, we briefly summarize the key results, first in nucleus--nucleus collisions and then in \ee and deep inelastic \eD scattering processes.

\subsection{CSV in nucleus--nucleus collisions}

\begin{figure}[h!]
    \centering
    \begin{minipage}{0.48\textwidth}
        \centering
\resizebox{\textwidth}{!}{
  \includegraphics{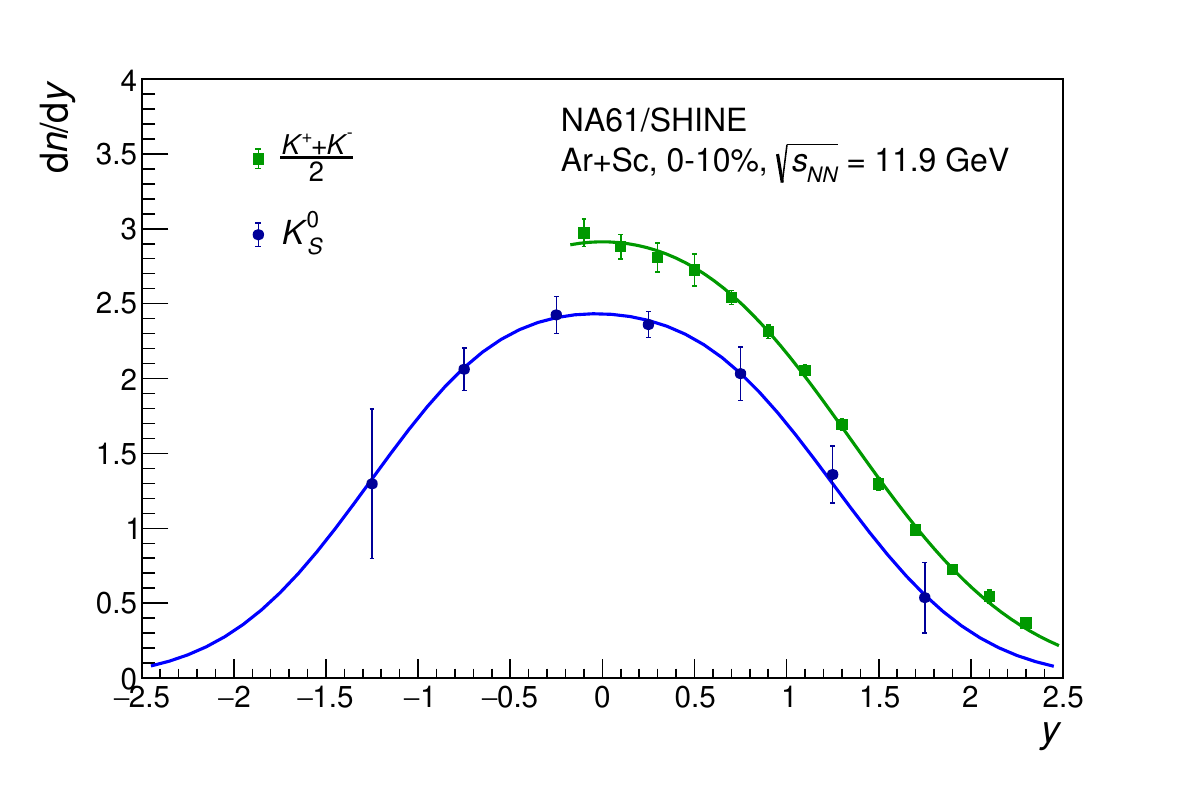} }
    \end{minipage} 
    \begin{minipage}{0.48\textwidth}
        \centering
        \vspace{1.2cm}
\resizebox{1.084\textwidth}{!}{
  \includegraphics{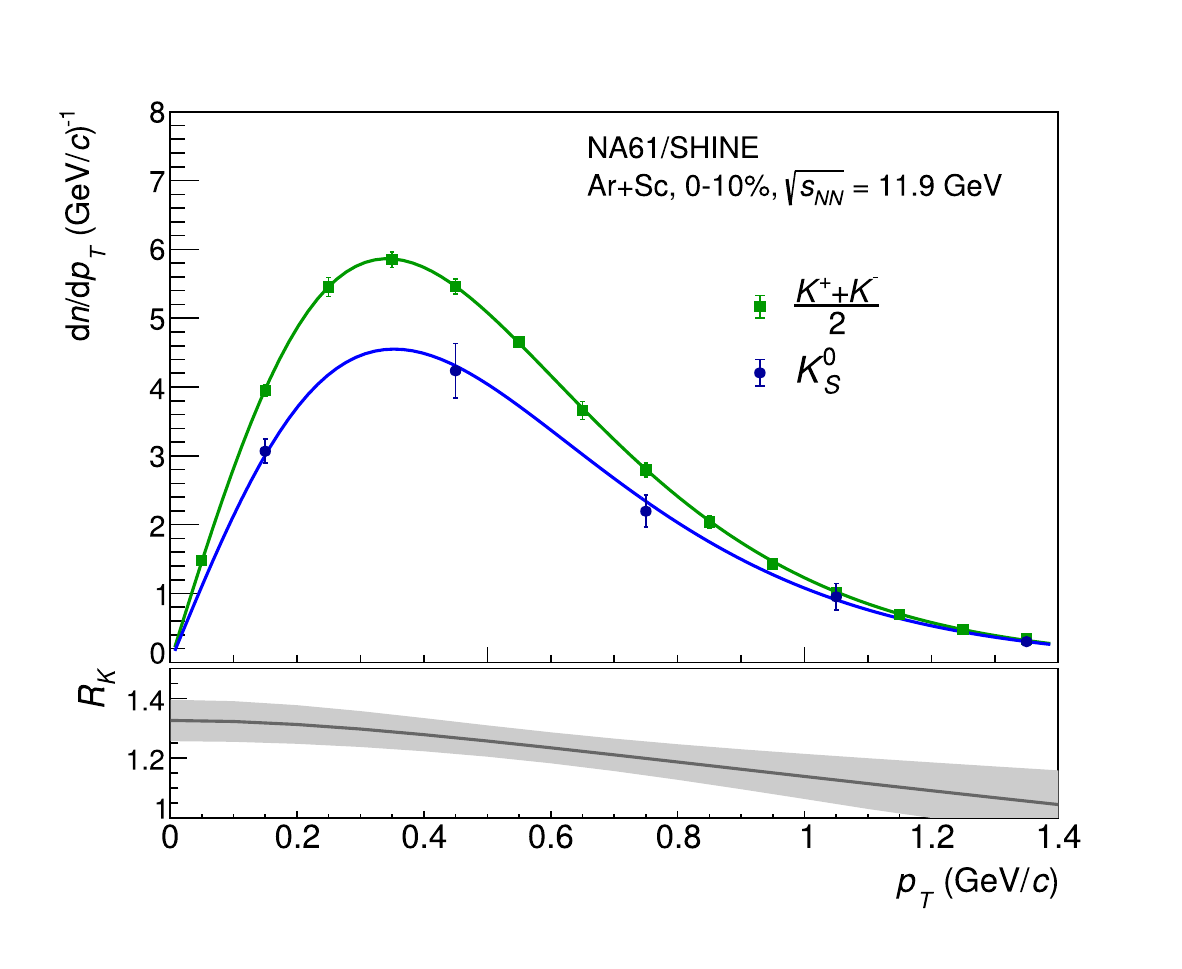} }
    \end{minipage}
    \vspace{-0.2cm}
    \caption{Comparison of rapidity (\textit{left}) and transverse momentum (\textit{right}) spectra of neutral (\ks) with the averaged spectrum of charged (($\kp+\km)/2$) kaons in the 10\% most central Ar+Sc collisions at 11.9~\GeV~\cite{NA61SHINE:2023azp}. Vertical bars denote total uncertainties.}
    \label{fig:NA61_kaons}
\end{figure}

Figure~\ref{fig:NA61_kaons} recalls the primary NA61/SHINE result: a comparison of charged and neutral kaon rapidity \y (\textit{left}) and transverse momentum \pt (\textit{right}) spectra in central Ar+Sc collisions at 11.9~\GeV~\cite{NA61SHINE:2023azp}.
An excess of charged kaons is visible at all rapidities. The charged-to-neutral kaon ratio decreases with increasing \pt,
reaching about 1.3 at \pt of several hundred~\MeVc and being consistent with unity at \pt $\approx$ 1.4~\GeVc.
Note that, 
the compilation of $R_K$ values shown in Fig.~\ref{fig:NA61_RK} also includes a new preliminary result for Ar+Sc collisions at 8.8~\GeV.

\begin{figure}[h!]
\centering
\resizebox{0.55\textwidth}{!}{
  \includegraphics{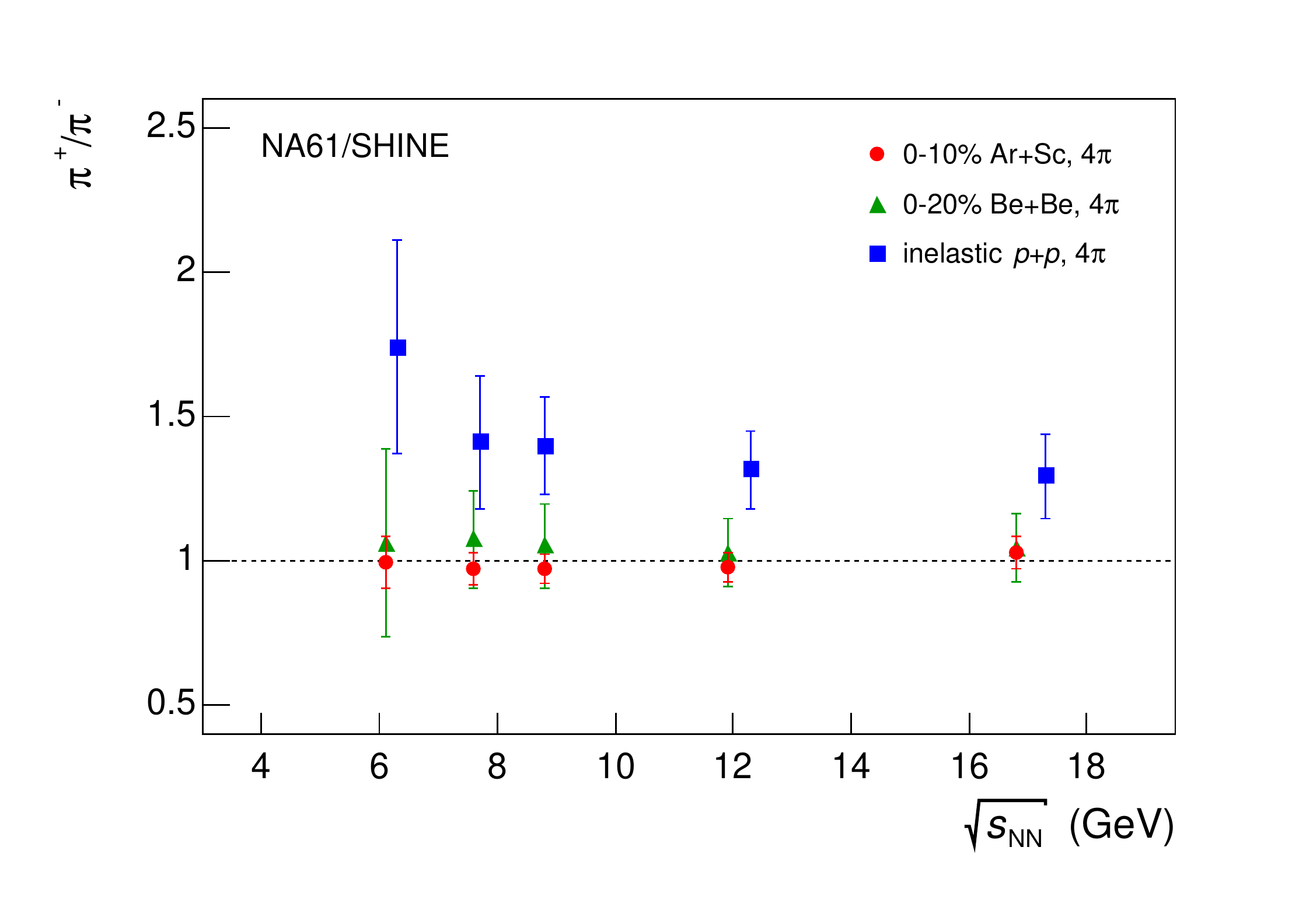}
}
\vspace{-0.5cm}
\caption{
Ratio of \pip to \pim yields in central Be+Be~\cite{NA61SHINE:2020czq} and Ar+Sc~\cite{NA61SHINE:2023epu} collisions as a function of collision energy as measured by NA61/SHINE. For comparison, results from the strongly charge-asymmetric \pp~\cite{NA61SHINE:2017fne} interactions are also shown. Vertical bars denote total uncertainties.
}
\label{fig:NA61_pions}
\end{figure}

NA61/SHINE reports no significant CSV in the pion sector. The ratio of \pip to \pim meson yields in central Be+Be and Ar+Sc collisions is consistent with unity within uncertainties; see Fig.~\ref{fig:NA61_pions}. 
Results on $\pi^0$ yield in nucleus--nucleus collisions 
are sparse~\cite{WA98:1998psk,PHENIX:2012jha,ALICE:2012wos,ALICE:2018mdl}. NA61/SHINE considers performing dedicated measurements that may improve the experimental situation~\cite{Pszczel_ISOBREAK25}.

NA49, the predecessor of NA61/SHINE, measured charged and neutral kaon yields in Pb+Pb collisions at the CERN SPS. The yields of charged kaons are published~\cite{NA49:2007stj,NA49:2002pzu}, whereas the \ks yields obtained in several analyses~\mbox{\cite{Margetis:1999ge,Barnby:thesis,Mischke:thesis,Strabel:thesis,Book:thesis}} remain preliminary (as summarized in Ref.~\cite{Stroebele_ISOBREAK25}).
This is due to unresolved inconsistencies in the \ks analysis and/or an observed suppression of the \ks yield relative to the charged kaon yield.
At the time, the \ks suppression was treated as an analysis-related inconsistency. The NA49 points shown in Fig.~\ref{fig:NA61_RK} are based on the published charged kaon results~\cite{NA49:2007stj,NA49:2002pzu} and preliminary \ks results~\cite{Strabel:thesis}.
The NA49 example suggests that other measurements related to the $R_K$ ratio may also be biased by the strong expectation that $R_K$ is close to 1.

\begin{figure}[h!]
\centering
\vspace{0.5cm}
\resizebox{0.75\textwidth}{!}{
  \includegraphics{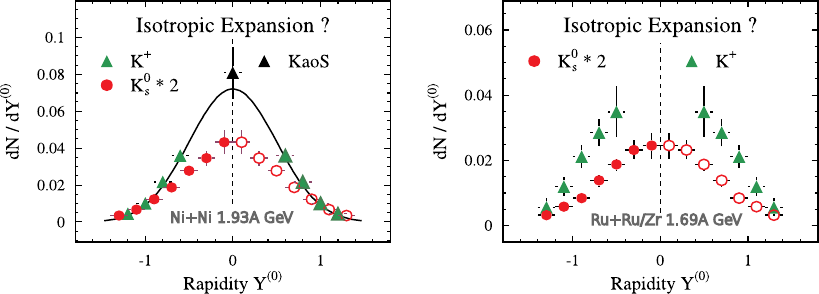}
}
\caption{
Results on charged and neutral kaon yields as a function of rapidity in nucleus--nucleus collisions from FOPI and KaoS at the GSI SIS18. Note that at these low collision energies, the \km yield is negligible. The plots are taken from Ref.~\cite{Kutsche:thesis} and presented in Ref.~\cite{Lorenz_ISOBREAK25}.}
\label{fig:SIS_RK}
\end{figure}

The low-energy end of the $R_K$ collision-energy dependence is set by measurements at the GSI SIS18. The most relevant point comes from HADES (see Fig.~\ref{fig:NA61_RK}). Figure~\ref{fig:SIS_RK} shows complementary information from the FOPI and KaoS experiments on rapidity distributions of charged and neutral kaons~\cite{Kutsche:thesis,Forster:2007qk}.
Note that at these low collision energies, the \km yield is negligible compared to the \kp and neutral kaon yields.
Measurements from different experiments at the GSI SIS18 establish a consistent picture of a significant excess of charged kaons spanning a broad rapidity range.

The STAR results~\cite{Zhang_ISOBREAK25} compiled in Fig.~\ref{fig:NA61_RK} show consistently the $R_K$ value of about 1.2, bringing together the SPS and LHC collision energy domains, but suffer from large uncertainties.

\begin{figure}[h!]
\centering
\begin{minipage}{0.4163\textwidth}
\centering
\includegraphics[width=\textwidth]{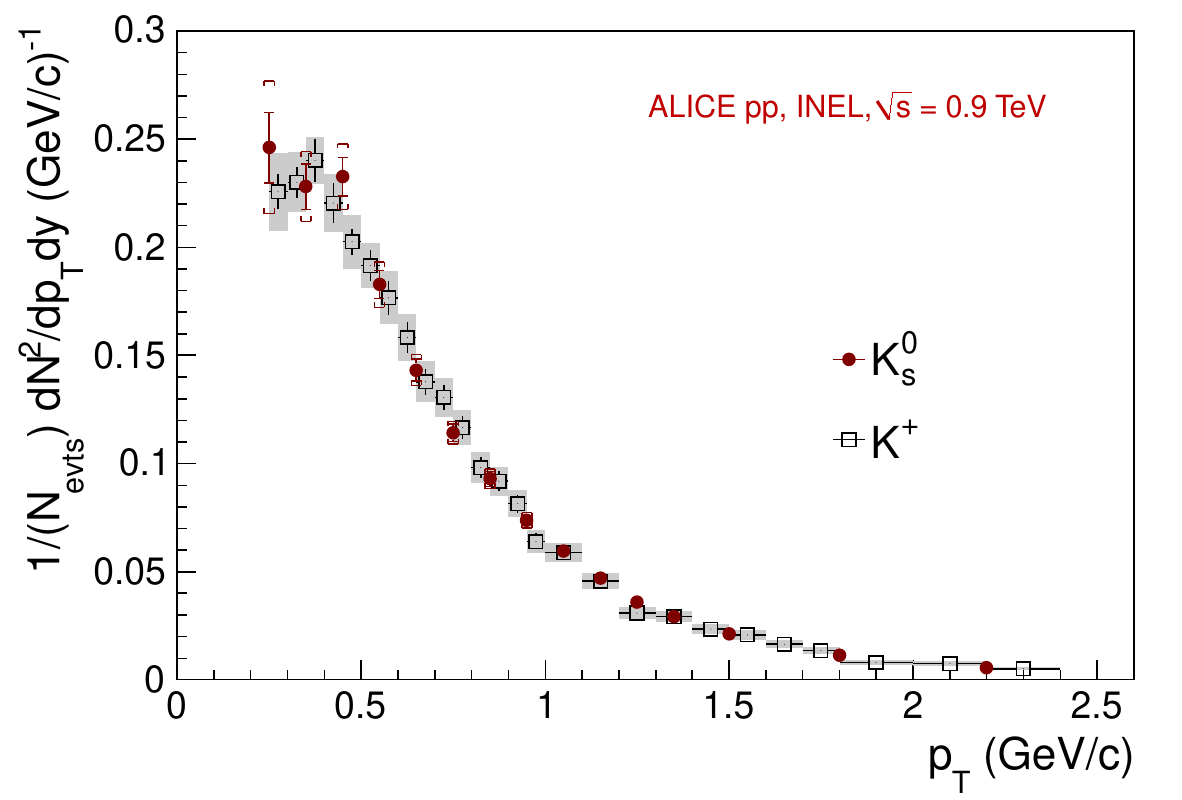}
\end{minipage}
\begin{minipage}{0.5745\textwidth}
\centering
\vspace{+0.305cm}
\includegraphics[width=\textwidth]{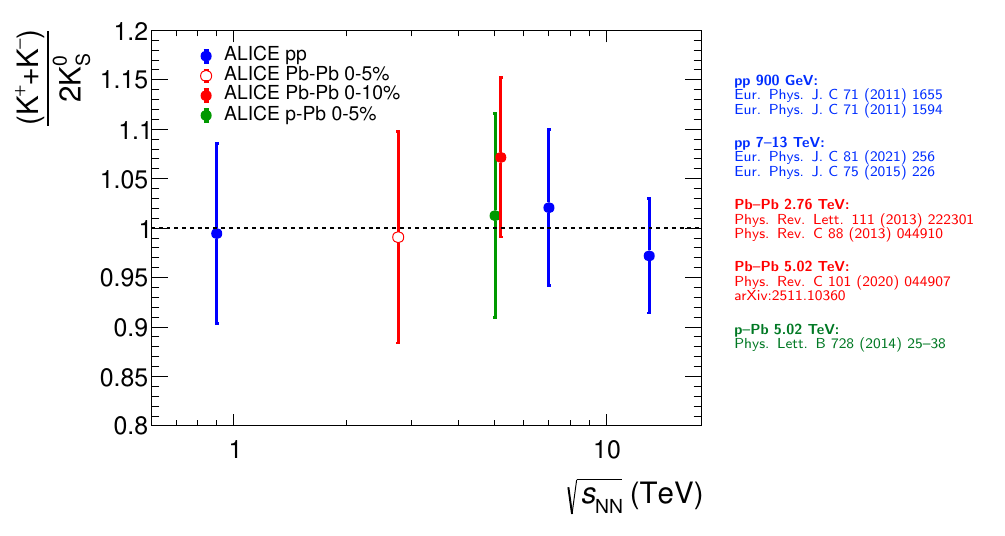}
\end{minipage}
\caption{
\textit{Left}: Comparison of transverse momentum spectra of \ks and \kp mesons in inelastic \pp interactions at 0.9~\TeV~\cite{ALICE:2010vtz}.  
\textit{Right}: The $R_K$ ratio in \pp, $p$+Pb, and Pb+Pb collisions at LHC energies as measured by ALICE. The plots are taken from Ref.~\cite{Ercolessi_ISOBREAK25}; the right one with later changes. 
}
\label{fig:ALICE_RK}
\end{figure}

Finally, the very high-energy end of the $R_K$ collision-energy dependence is provided by measurements from ALICE at the CERN LHC~\mbox{\cite{ALICE:2013mez,ALICE:2013cdo,Ercolessi_ISOBREAK25}}.
At LHC energies, particle and antiparticle yields at mid-rapidity are approximately equal, independent of the collision system. Thus, the yields of \kp and \ks mesons should be equal if charge symmetry holds.
This is indeed observed in Fig.~\ref{fig:ALICE_RK} (\textit{left}), where the \pt spectra of the two species are compared.
Figure~\ref{fig:ALICE_RK} (\textit{right}) shows the corresponding $R_K$ ratio measured in \pp, $p$+Pb, and Pb+Pb collisions; in all cases it is consistent with unity within about 10\% uncertainties.

In summary, the discrepancy between the initial experimental data and the HRG model was estimated in Ref.~\cite{NA61SHINE:2023azp} to have a significance of about $4.7\,\sigma$, assuming that the statistical and systematic uncertainties of the experimental points are mutually independent. As a cross-check, using the HRG implementation of Ref.~\cite{Brylinski:2023nrb}, we obtain a significance of $4.9\,\sigma$. For both HRG implementations, the value $\chi^2_{\rm min}/{\rm dof}\approx0.3$ may indicate either an overestimation of the experimental uncertainties or the presence of correlations among the measurements.

Repeating the analysis using all data points shown in Fig.~\ref{fig:NA61_RK}, including the new results from NA61/SHINE and ALICE, yields a data-to-HRG ratio of $1.116 \pm 0.022$ for the HRG implementation of Ref.~\cite{Brylinski:2023nrb}. Consequently, the significance of the discrepancy increases to $5.3\,\sigma$. The corresponding value of $\chi^2_{\rm min}/{\rm dof}\approx0.57$ remains significantly below unity.

To estimate the potential impact of correlations among measurements from the same experiment, we adopt the extreme assumption that the quoted uncertainties are entirely dominated by systematic effects and that these systematic uncertainties are fully correlated within each experiment. We therefore replace the individual measurements from each experiment with their arithmetic mean and assign to it the arithmetic mean of the corresponding uncertainties. The resulting experimental averages are then compared with the corresponding averages of the HRG predictions. This procedure leaves seven independent data points and yields a data-to-HRG ratio of about $1.113 \pm 0.036$, with $\chi^2_{\rm min}/{\rm dof}\approx0.94$ being close to unity.
This corresponds to a discrepancy with a significance of $3.2\,\sigma$ (based on precise numerical values).
At the opposite extreme, if the value $\chi^2_{\rm min}/{\rm dof}\approx0.57$ resulted entirely from an overestimation of the experimental uncertainties rather than from correlations, rescaling the uncertainties to obtain $\chi^2_{\rm min}/{\rm dof}=1$ would increase the significance to $5.3/\sqrt{0.57}\approx7\,\sigma$.
Thus, the significance of the discrepancy lies in the range $(3-7)\sigma$, depending on the assumptions about correlations among measurements. 
Finally, we note that, as suggested by the review of the NA49 results above, the central $R_K$ values may also be unconsciously underestimated by the strong expectation that $R_K$ is close to 1. This could lead to a preference for considering and correcting biases that increase $R_K$ rather than those that decrease it; see Ref.~\cite{ParticleDataGroup:2024cfk} for historical examples of the time dependence of key physics quantities.

A review of the model predictions for $R_K$, providing input to theoretical uncertainties, is provided in Sec.~\ref{sec:legacy}.

\subsection{$\pmb{e^+}$+$\pmb{e^-}$ interactions and deep $\pmb{e^-}$+D inelastic scattering}

\begin{figure}[h!]
\centering
\resizebox{0.85\textwidth}{!}{
  \includegraphics{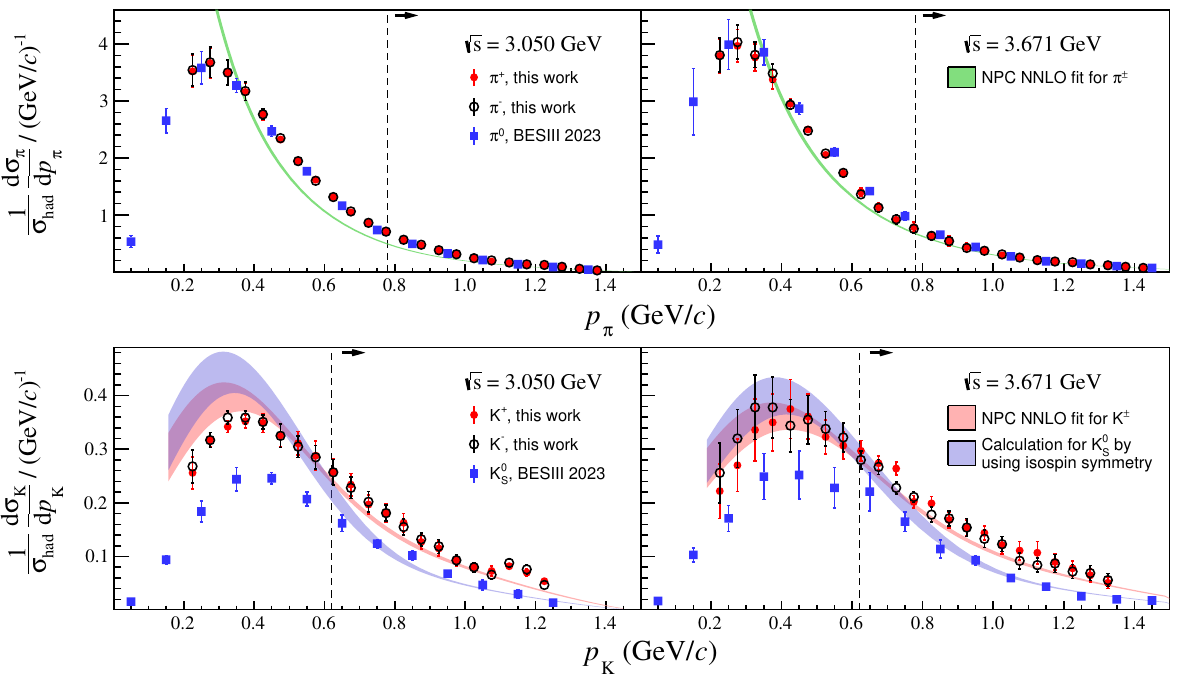}
}
\caption{
Normalized differential cross sections of charged and neutral pions (\textit{top}) 
and kaons (\textit{bottom}) as a function of hadron momentum in \ee interactions at 3.050 and 3.671~\GeV. Vertical
bars denote total uncertainties.
The green and red bands denote the ``NPC'' (Nonperturbative Physics Collaboration) NNLO calculations~\cite{Gao:2024dbv,Gao:2024nkz} with $1\sigma$ limits, based on a new global fit including world data~\cite{Gao:2024dbv} and the new BESIII results.
The blue band represents the ``NPC'' NNLO calculation for \ks using $K^\pm$ fragmentation functions via isospin symmetry.
Only the measurements to the right of the dashed lines are used in the fit. 
The plots are taken from Ref.~\cite{BESIII:2025mbc} and presented in Ref.~\cite{Huang_ISOBREAK25}.
}
\label{fig:BESIII}
\end{figure}

Figure~\ref{fig:BESIII} compares momentum (\p) spectra of pions (\textit{top}) and kaons (\textit{bottom}) produced in \ee interactions at 3.050 and 3.671~\GeV~\cite{BESIII:2025mbc}.
The \pip, $\pi^0$, and \pim spectra agree well with each other.
This is not the case for kaons: the charged kaon yields are, on average, about 1.5 times higher than those of \ks.
The excess is observed in the measured momentum range 0.2--1.5~\GeVc.
A model fitted to the $K^\pm$ spectra assuming QCD charge symmetry and its QED violation describes the \ks results well down to \p $\approx$ 0.6~\GeVc but significantly overpredicts the neutral kaon yield at lower momenta.
This behavior may indicate CSV in quark fragmentation functions.

\begin{figure}[h!]
\centering
\resizebox{0.6\textwidth}{!}{
  \includegraphics{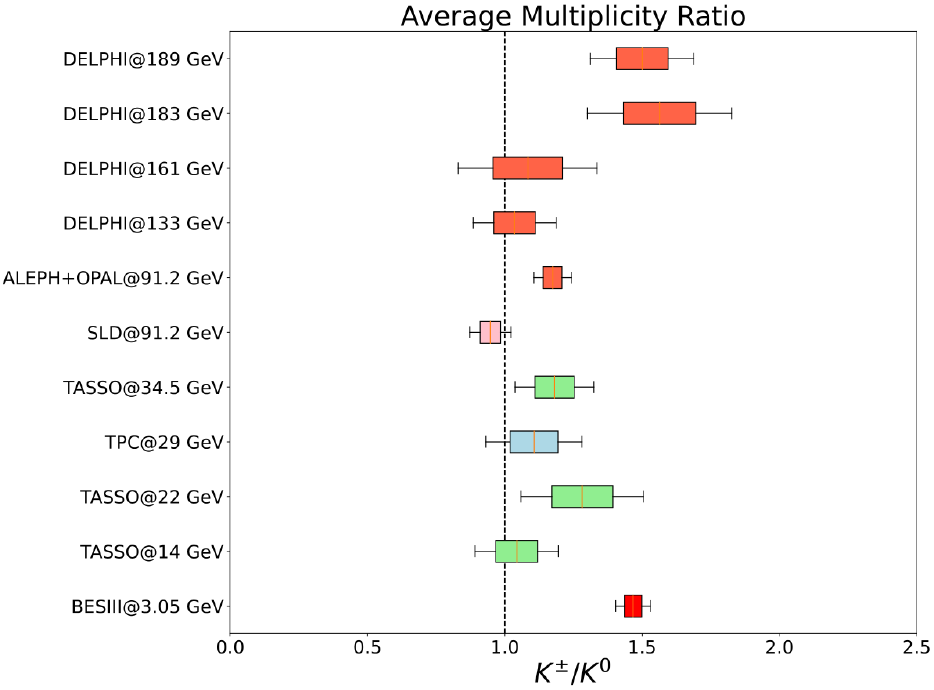}
}
\caption{
Compilation of world data on the charged-to-neutral kaon ratio in \ee interactions. 
The plot is taken from Ref.~\cite{Huang_ISOBREAK25}. 
}
\label{fig:e+e-}
\end{figure}

Figure~\ref{fig:e+e-} presents a compilation of the charged-to-neutral kaon ratio in \ee interactions measured by various experiments over the broad energy range 3--200~\GeV.
There is tension among the results, with $R_K$ ranging from approximately 1 to approximately 1.5. The world average is about 1.2. A detailed analysis of the compiled data to clarify the origin of the tension is important.

\begin{figure}[h!]
\centering
\resizebox{0.55\textwidth}{!}{
  \includegraphics{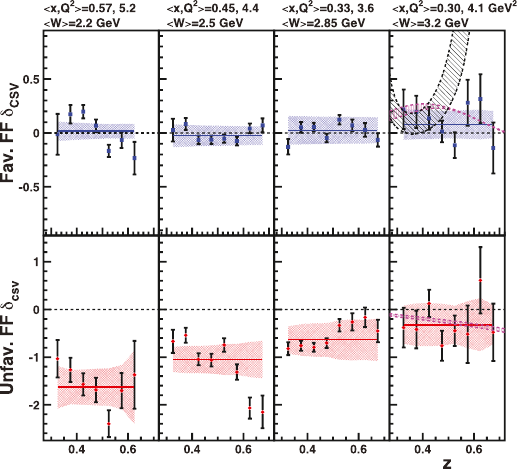}
}
\caption{
The $z$ (pion's longitudinal momentum fraction) dependence of the charge-symmetry-violating parameter $\delta_{\mathrm{CSV}}$ for the favored (dominant) fragmentation function (\textit{top} panels) and the unfavored (subdominant) fragmentation function (\textit{bottom} panels), extracted from measured charged pion multiplicities on hydrogen and deuterium targets. The panels are ordered in decreasing $x$ (increasing $W$). Here $x$, $W$, and $Q^2$ denote the nucleon momentum fraction, center-of-mass energy, and square of the four-momentum transfer, respectively. The blue (red) solid lines show constant fits to the favored (unfavored) $\delta_{\mathrm{CSV}}$. Under charge symmetry, $\delta_{\mathrm{CSV}}=0$, as indicated by the black dashed lines. For details, see Ref.~\cite{Bhatt:2024prq}. The plots are taken from Ref.~\cite{Bhatt:2024prq} and presented in Ref.~\cite{Dutta_ISOBREAK25}.
}
\label{fig:SIDIS}
\end{figure}

Independently, Jefferson Lab measured the flavor dependence of pion multiplicities in semi-inclusive deep inelastic scattering (SIDIS) of 10.2 and 10.6~\GeV electrons on proton and deuteron targets to investigate possible CSV in fragmentation functions.
Assuming factorization at low \pt and allowing for CSV, the data can be described by two ``favored'' and two ``unfavored'' flavor-dependent effective low-\pt fragmentation functions.
Figure~\ref{fig:SIDIS} summarizes the extracted CSV parameters~\cite{Bhatt:2024prq}.
The favored fragmentation functions are flavor independent over the covered center-of-mass energy ($W$) range, while the unfavored fragmentation functions exhibit an increasing flavor dependence toward lower $W$.
The sum and difference ratios of deuteron-to-hydrogen multiplicities also show CSV at low $W$ but gradually approach the charge-symmetry expectation with increasing $W$.
These results may indicate CSV in pion fragmentation, although contributions from higher-twist effects cannot be excluded. Planned global fits of \ee and DIS data will help distinguish between these interpretations.

Thus, results from \ee annihilation and deep inelastic processes point to the same qualitative conclusion: quark fragmentation functions seem to violate charge symmetry.

%% file: sections/Th_legacy.tex

\begin{figure}[h!]
\centering
\resizebox{0.5\textwidth}{!}{
  \includegraphics{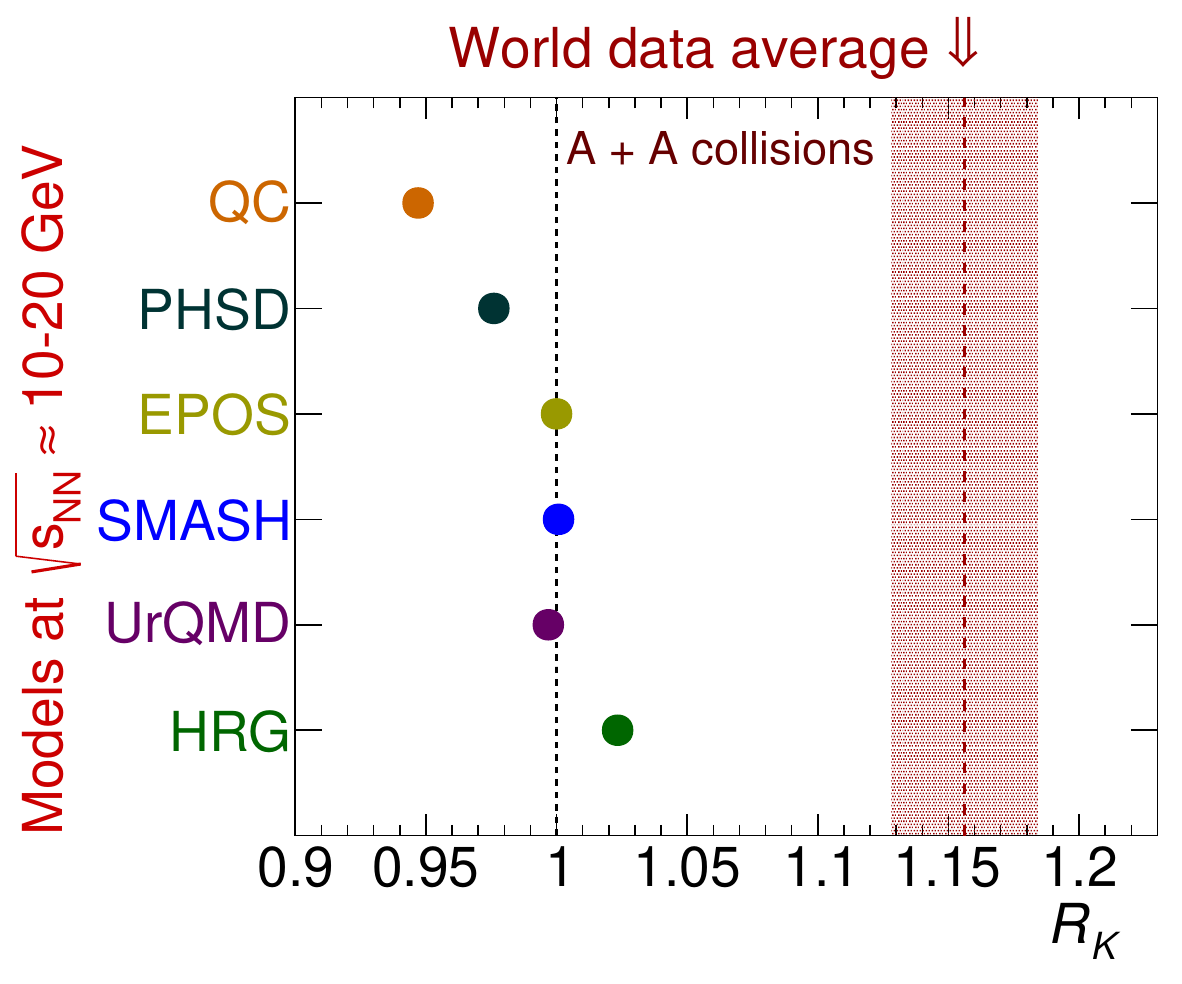}
}
\caption{
Compilation of model predictions for $R_K$ in nucleus--nucleus collisions in the range \snn = 10--20~\GeV for $Q/B=0.4$, close to the values taking place in the experiments listed in  Fig. \ref{fig:NA61_RK}. The world data average includes all data points for \mbox{\snn $\geq$~2.6~\GeV} as in Table~3 of Ref.~\cite{NA61SHINE:2023azp}. See text for further details.
}
\label{fig:models_compil}
\end{figure}

Here, we briefly summarize predictions for the $R_K$ ratio from widely used (``legacy'') models of nucleus--nucleus collisions. The results are compiled in Fig.~\ref{fig:models_compil}. Despite substantial differences in their physics assumptions, all models predict $R_K$ values close to unity, missing CSV observed in the data.
Note, the theoretical results correspond to $Q/B=0.4$, close to the values realized in the experiments listed in Fig. \ref{fig:NA61_RK}. 
Using, instead, $Q/B=0.5$ modifies the result for the predicted $R_K$ only slightly.

The individual model results shown in Fig.~\ref{fig:models_compil} are detailed below.

\begin{itemize}

\item \textbf{Hadron Resonance Gas (HRG)}~\cite{Kapusta_ISOBREAK25,Gorenstein_ISOBREAK25} assumes statistical equilibrium at chemical freeze-out. It incorporates known isospin-breaking effects implicitly, such as mass splittings within isospin multiplets and different branching ratios into isospin-related channels (e.g.\ $\phi(1020)\rightarrow \kp \km$ vs.\ $\phi(1020)\rightarrow K^0\overline{K}^0$). 
For $Q/B = 0.4$ at $\snn = 20$~\GeV, the model yields~\cite{NA61SHINE:2023azp,Samanta_ISOBREAK25} (as an example of earlier analysis see Ref.~\cite{Letessier:2005qe}):
\[
R_K \simeq 1.023\, .
\]
The quoted value does not include the uncertainty that arises from different chemical freeze-out parameterizations and resonance lists.
In Ref.~\cite{Brylinski:2023nrb}, the energy dependence of $R_K$ is presented for both $Q/B = 0.4$ and $Q/B = 0.5$. For \snn $>$ 10~\GeV, the corresponding predictions differ only marginally and remain close to unity. 

\item \textbf{Ultra-relativistic Quantum Molecular Dynamics (UrQMD)}~\cite{Bleicher:2022kcu,Bleicher:1999xi} is a microscopic hadronic transport model based on covariant propagation and resonance and string dynamics. For Au+Au collisions at $\snn = 20$~\GeV, it predicts~\cite{NA61SHINE:2023azp}
\[
R_K = 0.9969 \pm 0.0027\, ,
\]
consistent with unity within a small statistical uncertainty.

\item \textbf{Parton--Hadron-String Dynamics (PHSD)}~\cite{Bratkovskaya:2011wp} is an off-shell transport model that includes partonic, hadronic, and string degrees of freedom with dynamical hadronization. For Ar+Sc collisions at $\snn =$ 11.9~\GeV (0--10\%), it gives 
\[
R_K \simeq 0.976\, ,
\]
with the numerical value obtained via private communication~\cite{Ozvenchuk_ISOBREAK25}.

\item \textbf{SMASH}~\cite{SMASH:2016zqf} is a hadronic transport model optimized for precision studies of the hadronic phase. For central Pb+Pb collisions (impact parameter $b=0$--3~fm) at $\snn =$ 11.9~\GeV at mid-rapidity, it yields  (with or without neutron skin)~\cite{Vitiuk_ISOBREAK25}
\[
R_K \simeq 1\, .
\]

\item \textbf{EPOS}~\cite{Pierog:2013ria} is a hybrid approach based on Gribov--Regge multiple scattering, string fragmentation, and collective effects. For Au+Au collisions at $\snn = 39~\GeV$, it predicts 
\[
R_K \simeq 1\, ,
\]
with no intrinsic asymmetry between charged and neutral kaons from string fragmentation~\cite{Bleicher_ISOBREAK25}. A similar result is expected at 20~\GeV.

\item \textbf{Quark Coalescence (QC)}~\cite{Bonesini:2001iz,Stepaniak:2023pvo} (and references therein) assumes that hadrons are formed via quark and antiquark coalescence. In the absence of explicit isospin breaking, one obtains
\[
R_K = 0.947 \pm 0.004
\]
for $Q/B=0.4$ at $\snn = 20$~\GeV (the numerical value from Ref.~\cite{Rohrmoser_private} is based on Refs.~\cite{Giacosa:2025ynn,Rohrmoser_ISOBREAK25}). 

\end{itemize}

\paragraph{Summary of legacy models.}

Despite their conceptual diversity -- ranging from equilibrium statistical models to microscopic hadronic transport and partonic approaches -- all tested legacy models predict $R_K \approx 1$, in disagreement with experimental results, see Fig.~\ref{fig:models_compil}. It is important to stress that these models, although inspired by QCD, are not derived directly from the QCD Lagrangian. 

\subsection*{Perspective from lattice QCD}

Lattice QCD~\cite{Luscher:2010ae,Philipsen:2012nu} provides first-principles calculations of strongly interacting matter in equilibrium, primarily at small baryon densities. However, it is currently far from capable of predicting hadron multiplicities in nucleus--nucleus collisions.

Isospin breaking due to $m_u \neq m_d$ has been implemented in lattice calculations~\cite{Frezzotti:2022dwn,deDivitiis:2011eh,BMW:2014pzb,Giusti:2017dmp,FlavourLatticeAveragingGroupFLAG:2024oxs}, successfully describing vacuum mass splittings within isospin multiplets. At finite temperature, however, only one exploratory study with $m_u \neq m_d$ exists so far~\cite{Gavai:2002fi}, and it finds no significant modification of the critical temperature, in agreement with isospin symmetry.

A systematic investigation of thermodynamic properties in lattice QCD at finite temperature with $m_u \neq m_d$ remains an open task. In the vacuum, the difference between the condensates $\langle u\overline{u} \rangle$ and $\langle d\overline{d} \rangle$ provides an additional measure of isospin breaking that can, in principle, be addressed within lattice QCD~\cite{Brandt_ISOBREAK25}. Whether such effects are negligible or -- quite surprisingly -- could dynamically generate an asymmetry large enough to explain the observed CSV in nucleus--nucleus collisions remains unknown.

%% file: sections/Th_fit.tex
Since widely used models fail to reproduce the experimental value of $R_K$, the most direct way to describe the data has been to introduce explicit charge-symmetry-violation parameters. While this strategy allows for quantitative agreement with measurements, it does not in itself explain the physical origin of the effect. Rather, it shifts the problem to the level of parametrization.

The UrQMD model was the first to implement such an explicit breaking of isospin symmetry~\cite{Reichert:2025znn}. In this approach, the string fragmentation mechanism is modified such that the production of $u\overline{u}$ pairs is assumed to be three times more probable than that of $d\overline{d}$ pairs:
\[
P(u\overline{u})/P(d\overline{d}) = 3\, .
\]
With this assumption, the measured $R_K$ values in \ee collisions at $\sqrt{s} = 3.050$~\GeV and in \pp collisions at $\snn = 12.3$~\GeV can be reproduced. Using the same parameter value, UrQMD also describes the NA61/SHINE kaon data in Ar+Sc collisions.

Within the SMASH framework, high-energy particle production is modeled via string excitation and fragmentation, implemented using \textsc{Pythia}~\cite{Bierlich:2022pfr}. A best fit to the Ar+Sc data at $\snn =$ 11.9~\GeV is obtained for (see Ref.~\cite{Vitiuk_ISOBREAK25}):
\[
P(u\overline{u})/P(d\overline{d}) = 1.89\, .
\]

In both UrQMD and SMASH, the charged kaon excess is introduced via a substantial enhancement of $u\overline{u}$ over $d\overline{d}$ production during string fragmentation. 
Notice also that the introduced string fragmentation asymmetry results in $R_K>1$ in the central rapidity region, independent of the initial state. This is because the string fragmentation used is universal and is independent of the strings' ends. For example, this implies $R_K >1$ for $\pi^{\pm}$+C interactions, see discussion in Sec.~\ref{sec:next} for the relevance of these reactions.

The quark coalescence model~\cite{Giacosa:2025ynn} provides a different parametrization of the observed CSV. There, the data on $R_K$ are fitted by assuming that the QCD vacuum may contain unequal number of $u\overline{u}$ and $d\overline{d}$ quark pairs. The fit to nucleus--nucleus data resulting in 20\% more $u\overline{u}$ than $d\overline{d}$ pairs, yields $R_K \simeq r \simeq1.2$ in both $\pi^+$+C and $\pi^-$+C collisions, where $r$ is the ratio of $u\overline{u}$ and $d\overline{d}$ emerging from the QCD vacuum. It also leads to other predictions, such as $p/n \simeq  R_K \simeq 1.2$, $\Sigma^+/\Sigma^- \simeq  R_K^2  \simeq 1.4$, and  $\Delta^{++}/\Delta^- \simeq  R_K^3  \simeq 1.7$  in nucleus--nucleus collisions at $\snn = 20$~\GeV. On the other hand, the $\pi^+ / \pi^0$ ratio is predicted to be close to unity -- it is only slightly dependent on the $r$ parameter~\cite{Giacosa:2025ynn}. The above predictions allow for experimental tests of the model. Still, an origin of the fitted $u\overline{u}$ and $d\overline{d}$ asymmetry remains unexplained.

A similar phenomenological extension could be introduced in statistical hadronization models by allowing separate fugacity factors for $u$ and $d$ quarks~\cite{Petran:2013lja,Rafelski:2014cqa,Rafelski:2015hta}. Such a modification would explicitly encode isospin asymmetry prior to hadronization. 
The SHARE~\cite{Torrieri:2004zz,Torrieri:2006xi,Petran:2013dva} model remains to this date the only statistical hadronization model which implements baryon and charge conservation and allows for a  $u\overline{u}$ versus $d\overline{d}$ pair asymmetry by the parameter $\gamma_3$. Event-generator implementation of the statistical hadronization framework such as THERMINATOR can incorporate similar effects~\cite{Ryblewski_ISOBREAK25}. This may allow for predictions within a given experimental acceptance.

In contrast, the HRG model -- being an equilibrium framework constrained by the hadron spectrum listed in the PDG \cite{ParticleDataGroup:2024cfk}
-- cannot be modified straightforwardly. As discussed in Refs.~\cite{NA61SHINE:2023azp,Brylinski:2023nrb}, known sources of isospin-symmetry breaking are already included, and additional contributions from resonances such as $a_0(980)$ and $f_0(980)$ were estimated to affect $R_K$ only marginally.

%% file: sections/Brain.tex
Here, we outline several possible physics effects that may contribute to an explanation of the difference between models and data on $R_K$. Below, we summarize ideas and open questions, some of which have emerged during discussions at ISO-BREAK~25.

\begin{enumerate}[(i)]

    \item \textbf{Effective charge-to-baryon ratio $\pmb{(Q/B)_\mathrm{eff}}$.}  
    In most model implementations, the charge-to-baryon ratio of the colliding nuclei ($Q/B$) is assumed to coincide with the effective ratio for the participating nucleons, $(Q/B)_{\mathrm{eff}}$. However, this assumption may be incorrect.
    The formal discussion of the selection rules due to isospin symmetry was discussed in Ref.~\cite{Turko_ISOBREAK25} (see also Ref.~\cite{Tinti_ISOBREAK25}).
\begin{itemize}
    \item The spatial distributions of protons and neutrons in heavy nuclei differ. In particular, the outer layers are neutron-rich~\cite{Novario:2021low}. In non-central collisions, this may reduce $(Q/B)_{\mathrm{eff}}$ relative to the ($Q/B$) value. This effect should, however, decrease $R_K$ and is possibly relevant only for very peripheral collisions.
    
    \item The baryon junction mechanism, which is expected to transport baryon number and enhance baryon stopping in heavy-ion collisions, may further modify the effective charge content at mid-rapidity~\cite{Kharzeev:1996sq,Pihan:2024lxw}. This could lead to a reduction in $(Q/B)_{\mathrm{eff}}$ and thus $R_K$ in the particle-production region.
    
    \item The sea-quark content of protons and neutrons is different. For example, the proton contains more $d\overline{d}$ than $u\overline{u}$ pairs~\cite{SeaQuest:2021zxb}, an effect that can be related to the Pauli principle~\cite{Geesaman:2018ixo} (see also Ref.~\cite{Ikeno:2019mne}). Neutrons exhibit the opposite trend. In neutron-rich systems ($Q/B < 1/2$), this may imply an excess of $u\overline{u}$ over $d\overline{d}$ pairs in the initial state, potentially influencing charged versus neutral kaon production for collisions with different proton and neutron compositions.

     \item  The $dd$ and $uu$ diquark binding energy may differ. If the former is more bound, the $u$ quarks are favored. This idea, however, is not favored because nucleons are predominantly composed of $ud$ diquarks~\mbox{\cite{Jaffe:2004ph,West:2020tyo}}. 
\end{itemize}
The impact of the above mechanisms on $R_K$ should diminish as $Q/B \to 1/2$ and vanish for collisions of fully symmetric light nuclei with the $\alpha$-cluster structure. This structure implies close-to-identical proton and neutron distributions. This motivates dedicated experimental measurements for collisions of light nuclei.

    \item \textbf{Electromagnetic effects.}  
    Although electromagnetic contributions are typically suppressed by $\alpha^2$, there may be specific situations in which they become non-negligible.
    \begin{itemize}
        \item At the hadronic level, the decays $a_0(980) \rightarrow \kp \km $ and $f_0(980) \rightarrow \kp \km$ near threshold receive non-perturbative Coulomb enhancements. The correction to the amplitude scales as $\alpha/v$, where $v$ is the relative velocity of the charged kaons and becomes small close to threshold. The resulting modification of the branching ratios is likely modest and confined to a narrow kinematic window~\cite{Giacosa_ISOBREAK25}, yet it may be worth quantifying it precisely.
        
        \item Strong transient electromagnetic fields generated in the early stages of relativistic heavy-ion collisions~\cite{Price:2023cll} could induce charge-dependent effects in hadron production~\cite{Rafelski_ISOBREAK25}. Whether such fields can produce a measurable difference between the yields of charged and neutral kaons remains an open question. For more details, see Appendix~\ref{appR}.
    \end{itemize}

    \item \textbf{Finite-density effects.}  
    Recent analyses~\cite{Stepaniak:2023pvo,Giacosa:2025ynn} (see also Refs.~\cite{Stepaniak_ISOBREAK25, Rohrmoser_ISOBREAK25}) indicate that \pp data may be consistent with obeying isospin symmetry. In contrast, nucleus--nucleus data -- apart from the ALICE measurements -- show clearly CSV.
    This may suggest importance of the large baryon density in the hadronic phase of nucleus--nucleus collisions at energies below those at the LHC. 
    Modifications to hadron properties, including kaons, have been widely discussed in the past; for a review, see Refs.~\cite{Rapp:1999ej,Ko:1997kb}. The effect may be different for charged and neutral kaons for collisions with $(Q/B)_{\mathrm{eff}} < 1/2$~\cite{Fuchs:2005zg}. While not specific to kaons, studies of QCD at finite isospin chemical potential may provide insight into dense matter~\cite{Ivanytskyi_ISOBREAK25,Ivanytskyi:2025cnn}.

    \item \textbf{Role of $\pmb{u}$--$\pmb{d}$ mass difference in $\pmb{q\overline{q}}$ creation.}  
     Assuming that $q\overline{q}$ pairs are mostly produced in the color octet state, when  $q$ and $\overline{q}$ repel each other, the $u\overline{u}$ quarks with small relative momenta are produced more abundantly than $d\overline{d}$ quarks. This happens because $d$ quarks are heavier than $u$ and the color Bohr radius of $d\overline{d}$ is smaller than that of $u\overline{u}$. Consequently, the production of $u\overline{u}$ is energetically more favorable than $d\overline{d}$~\cite{Mrowczynski_ISOBREAK25}. See Appendix~\ref{appM} for details. 

    \item \textbf{Role of the chiral anomaly.}  
    The chiral anomaly plays an important role in the neutral pseudoscalar meson sector. For the physical vacuum parameters, isospin-breaking effects 
    induced by $m_u \neq m_d$ remain small. Namely,  large effects associated with
    ($\pi^{0}$--$\eta$--$\eta'$) mixing would arise only in an extreme, fine-tuned limit in 
    which the anomalous contribution is almost completely suppressed~\cite{Gross:1979ur}\footnote{We note, however, that the schematic treatment of the $\eta'\to3\pi$ decay in terms of $\pi^0$--$\eta$ mixing proposed in Ref.~\cite{Gross:1979ur} was subsequently shown to be insufficient: direct isospin-breaking contributions and final-state interactions also play an important role~\cite{Borasoy:2006uv}.}. However, if the anomalous $U(1)_A$ contribution is strongly suppressed, as may occur toward effective $U(1)_A$ restoration at nonzero temperature, the resulting changes in the $\pi^0$--$\eta$--$\eta'$ system could enhance isospin-breaking effects \cite{Pisarski:1983ms}.  Moreover, because kaons do not couple directly to the anomalous flavor-singlet channel, any resulting effect on the charged-to-neutral kaon ratio would be indirect~\cite{Giacosa:2023fdz,Pisarski_ISOBREAK25}. Although speculative, exploring such indirect connections may provide additional insight.
    
    In fact, within nuclear collisions, the gas of hadrons is realized at nonzero temperature and density. The chiral anomaly may get partially restored in the medium, see e.g. Refs.~\cite{Tomiya:2016jwr,Aarts:2026kpq,Nicola:2020smo,Braun:2020mhk,Vertesi:2009ca} and references therein. Hence, this partial restoration may also affect isospin-sensitive quantities (even if it does not by itself imply a strong quantitative effect). The specific realization of the anomaly in effective models may also influence the order of the phase transition~\cite{Pisarski:1983ms,Giacosa:2024orp}. 

    \item \textbf{Disoriented-Chiral-Condensate.}
    Creating Disoriented-Chiral-Condensate (DCC) domains in heavy-ion collisions has been considered for many years~\cite{Anselm:1991pi,Blaizot:1992at,Rajagopal:1993ah}.
    They may be signaled by large fluctuations of the charged-to-neutral pion~\cite{Bjorken:1993wj} and kaon ratios~\cite{Schaffner-Bielich:1998mra,Gavin:2001uk}. A puzzling result on kaon fluctuations was recently reported by ALICE at LHC~\cite{ALICE:2021fpb}.
    Its possible interpretation by the DCC or by the formation of disoriented isospin condensates is discussed in Refs.~\cite{Kapusta:2022ovq,Kapusta:2023xrw}. The considered models for the charge-symmetric ensemble of collisions predict $R_K = 1$~\cite{SQM24}. The inclusion of isospin-breaking effects in the extended Linear Sigma model was recently discussed in 
    Ref.~\cite{Kovacs:2024cdb}, where through a fit to available experimental masses and decays of light mesons, it is shown that the relative difference between the $u\overline{u}$ and $d\overline{d}$ chiral condensates amounts to $0.02 \% $, implying that only very small deviations from $R_K =1$ are expected from this effect.
    
\end{enumerate}


%% file: sections/Next.tex
In spite of the challenges associated with the measurements and the limitations in our understanding of nuclear dynamics, we consider the current tension between theoretical predictions and experimental data sufficiently significant to warrant further investigation.
The question \emph{``what next?''} naturally concerns both experiment and theory. Experimental progress is essential for constraining theoretical models and identifying genuine physical mechanisms. At the same time, theory must evolve to provide a quantitative and internally consistent description of the relevant physics.

\subsection*{Experimental directions}

With the indications of CSV also in \ee and DIS processes, the central question has arguably shifted from \emph{``Where does the effect appear?''} to \emph{``Where does it not appear?''} This motivates a high-precision and systematic survey across reaction systems, collision centrality, and energies. For example, it is essential to clarify whether CSV persists at the highest (LHC) collision energies and whether it is present in pion-induced reactions. Such comparative studies will help narrow down the underlying mechanism.

The most immediate experimental steps include:

\begin{enumerate}[(i)]

\item Reducing experimental uncertainties. \\
A recurring recommendation was to adopt blind analysis techniques in order to minimize human bias in the results (see Ref.~\cite{Drachenberg_ISOBREAK25} for a review).

In contemporary measurements, systematic uncertainties dominate over statistical ones. Therefore, improving their estimation is of central importance. The application of the Barlow method~\cite{Barlow:2002yb} was suggested as a useful tool for a more robust treatment of systematics. In addition, experiments should clearly separate correlated and uncorrelated components of systematic uncertainties and, ideally, move toward unified procedures for their determination.

To this end, a working group with representatives from HADES, NA61/SHINE, and ALICE has been established to coordinate and harmonize these efforts.

\item Extending measurements to new systems and energies. \\
A systematic mapping of charge-symmetry violation across reaction systems and collision energies is crucial. 

NA61/SHINE has already recorded pion--carbon ($\pi^+$+$^{12}_6$C and $\pi^-$+$^{12}_6$C) and oxygen--oxygen ($^{16}_8$O+$^{16}_8$O) collisions~\cite{Kowalski_ISOBREAK25} (see also Ref.~\cite{Kowalski:2907307}). Further measurements with charge-symmetric initial ensembles are planned~\cite{Mackowiak-Pawlowska:2867952}. These data will be particularly valuable for isolating effects related to the effective $Q/B$ ratio.

HADES has collected and published $\pi^-$+$^{12}$C and $\pi^-$+$^{184}$W data at $\sqrt{s} = 2.02$~\GeV~\cite{HADES:2018qkj,HADES:2023sre}. The charged-to-neutral kaon ratios, integrated over \pt within the rapidity interval defined by the \kp data, are
\begin{equation*}
R_{\mathrm{C}} = 0.624 \pm 0.007\,(\mathrm{stat}) \pm 0.090\,(\mathrm{syst})\, ,
\end{equation*}
\begin{equation*}
R_{\mathrm{W}} = 0.738 \pm 0.005\,(\mathrm{stat}) \pm 0.130\,(\mathrm{syst})\, ,
\end{equation*}
as reported in Ref.~\cite{Lorenz_ISOBREAK25}. Because of the intrinsic charge asymmetry of the initial state, dedicated transport-model calculations (e.g.\ UrQMD) are required for quantitative interpretation. A dedicated analysis of the ratio \kp/\ks in a restricted phase space, designed to minimize systematic uncertainties, is currently in preparation.

Concerning results on DIS, the spectator-tagging technique pioneered at Jefferson Lab offers the possibility of accessing nearly free neutron targets, thereby improving constraints on unfavored fragmentation functions and their CSV~\cite{Bhatt:2024prq}.

Further proposals aim to use the 6--12~\GeV tagged photon beam in Hall~D at Jefferson Lab to measure charged-to-neutral pion and kaon ratios in photoproduction from deuterium and helium. This corresponds to a center-of-mass energy range $\sqrt{s} \approx$ 3.6--5~\GeV. These measurements would provide the first dedicated test of CSV in photonuclear reactions and would complement existing hadronic and \ee studies, helping to assess the universality of the observed effect.

\end{enumerate}

\subsection*{Testing hypotheses}

The experimental program outlined above will directly test several hypotheses discussed in Sec.~\ref{sec:brain}. For example, central O+O collisions involve small nuclei with nearly symmetric proton--neutron distributions, implying
\[
Q/B = (Q/B)_{\mathrm{eff}} = 1/2\, .
\]
If $R_K$ approaches unity in this case, it would strongly suggest that the previously observed excess of charged kaons originates from the neutron--proton asymmetry in heavier systems. 
Conversely, if $R_K \simeq 1.2$ persists even in such charge-symmetric systems, an entire class of possible explanations would be ruled out. Attention would then need to focus on mechanisms capable of generating an excess of charged over neutral kaons in a charge-symmetric environment. 

\subsection*{Theoretical challenges}

Current theoretical approaches require a significant $ud$-flavor symmetry violation to reproduce $R_K \simeq 1.2$~\mbox{\cite{Reichert:2025znn,Giacosa:2025ynn,Vitiuk_ISOBREAK25,Rafelski_ISOBREAK25}}. Without such a mechanism, $R_K$ remains close to unity. Understanding its origin constitutes a fundamental open question.
Improved theoretical modeling -- including refined effective approaches and lattice QCD calculations -- should aim to consistently incorporate all known sources of CSV, including electromagnetic effects.
Finally, extending the study to hadrons other than kaons should be considered part of the joint experimental and theoretical agenda.

%% file: sections/appR.tex
We consider~\cite{Rafelski_ISOBREAK25} the question of whether, in the early stage of quark--gluon plasma (QGP) formation, the mechanisms of QCD string breaking leading to the creation of abundant quark--antiquark pairs and entropy could be modulated by the strong electromagnetic (EM) field generated by relativistically moving large nuclear EM-charges. The initial asymmetry is almost certain to be large, as the generated collective magnetic field is large. However, in the evolving QGP phase, there are ongoing re-equilibration thermal processes such as $q\overline q\to q'\overline q'$, and any initial large asymmetry is going to be erased in dependence on the lifespan of the QGP phase and the evolution of the magnetic field in time. 
Therefore, it is necessary for isospin breaking to remain present that the strong EM field is preserved during the collision, assuring that the asymmetry is maintained. As we shall show, the state-of-the-art study of magnetic field evolution favors the NA61/SHINE collision-energy domain for further exploration of isospin asymmetry.   

Strong electromagnetic fields differentiate  $u$ and $d$ quarks, and this difference is, of course, what in the final state could cause the non-unity ratio $R_K$. 
We recall that the \(u\overline{u}\) pairs, with quarks carrying electric charge \(\pm2/3\), couple more strongly via the electromagnetic interaction than the \(d\overline{d}\) pairs, with charge \(\pm1/3\). In particular, since the coupling strength scales with the square of the charge, the \(u\overline{u}\) pair couples by a factor of four more strongly than the \(d\overline{d}\) pair.
In the presence of magnetic fields, the magnitude of the magnetic moment matters as well, $\mu=Gq/2mc$. 
Further, we recall that in addition to a factor $2$ in charge, there is also the mass ratio $m_u/m_d=0.460$, which implies that the magnetic moment coupling of perturbative $u$ quarks to the magnetic field is (assuming the same $ G=2$ factor) a factor $2/0.460=4.35$ times stronger compared to the $d$ quarks. Clearly, the in-matter magnetic moment could be modified considerably, as there is no conservation law that protects it; this medium effect is usually described as a modification of the $G$-factor. To conclude, it is self-evident that the lighter, twice-charged $u$ quarks will be much more affected by strong EM fields if present at the time of quark--gluon plasma formation. 

\begin{figure}
\centering
\vspace{3mm}
\includegraphics[width=8.5cm]{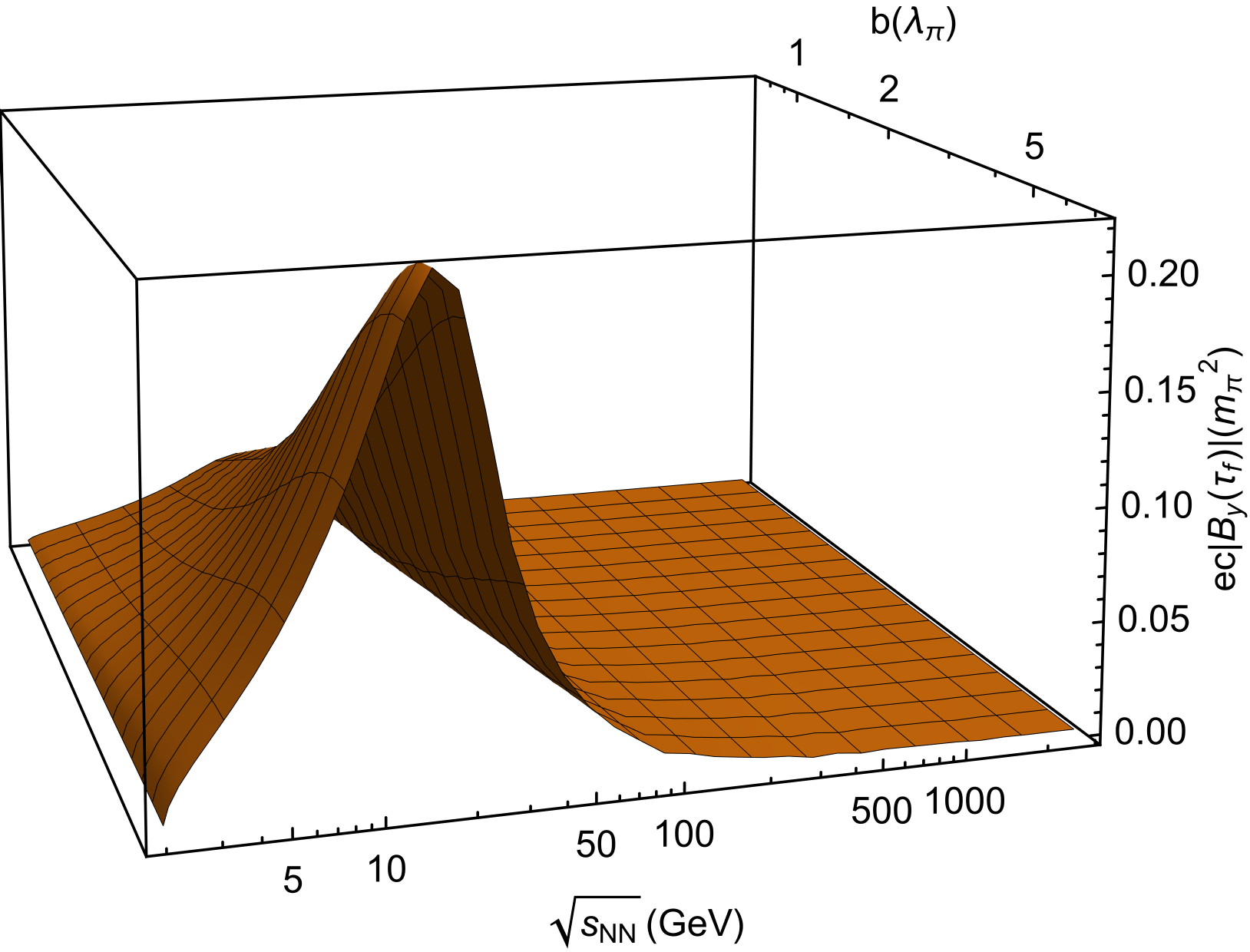} 
 \caption{The magnetic field in heavy-ion collisions as a function of impact parameter and collision energy at time of hadronization, the result courtesy of Chris Grayson, see also Ref.~\cite{Grayson:2022asf}.}
\label{magby}
\end{figure}

The EM fields of interest to us would need to compete with QCD parton formation mechanisms in strength. In general, the electric field is weaker than the magnetic field due to the mobility of neutralizing charges in the QGP.  Moreover, the magnetic field strength in relativistic ion collisions is at its maximum early on. As charges separate, this field diminishes. The magnitude of
the resulting residual field at hadronization is available from a fully dynamical calculation~\cite{Grayson:2022asf} and shown in
Fig.~\ref{magby}, where the magnetic field strength normal to the collision axis is given in units of $m_\pi^2$ as a function of \snn, with the impact parameter shown in units of $1/m_\pi$.  The most interesting aspect of this calculation was the strong maximum of the field, achieved at modest collision energies that are within the range available for exploration by the NA61/SHINE Collaboration. While our theory may evolve further, the possible appearance of a magnetic peak with a significant magnetic field across both impact parameter and collision energy is just what is needed to address the isospin asymmetry observed today.
 
The second point we address is the mechanism of interference between the QCD string breaking and this global magnetic field. In many kinetic models of QGP formation, the presumption is made that pair production rate is described by the imaginary part of an effective action first derived by Euler and Heisenberg, and hence called Schwinger 
pair production. 
We generalize for now considering the well-known total vacuum decay rate
to allow for both QCD string field strength with magnitude $a_c$ and with QCD coupling $g$, as well as for (electro)magnetic field strength, which we call $b_e$ with EM coupling $q$, and consider 
an arbitrary value of the $G$-factor presumed to be effectively the same for QCD and QED interaction: 
\begin{equation}
2 \operatorname{Im}V_{\rm eff}
   =\frac{g^{2}a^2_c}{4\pi^{3}}
     \sum_{n=1}^{\infty}\frac{ e^{-n\pi m^{2}_i/ga_c}}{n^2}\,
      \,\frac{n\pi q_ib_e/ga_c}{\sinh\!\bigl(n\pi q_ib_e/ga_c\bigr)}
     \cos\!\Bigl(\left(\tfrac{G}{2}+1\right)\,n\pi\Bigr)\,
      \cosh\!\Bigl(\tfrac{G}{2}\,\frac{n\pi q_ib_e}{ga_c}\Bigr)\, , 
      \label{imveff}
\end{equation}
where magnitudes of electric charge enter as $q_u=|2e/3|, q_d=|-1/3e|$ and the couplings as
\begin{equation}
   \tfrac{g^2}{4\pi} = \alpha_s ,\qquad \tfrac{e^2}{4\pi}=\alpha_{em} \text{ ,}  \qquad -2 \le G\le 2 
 ~~~~~\textrm{  (see text for a wider range)~.} 
\end{equation}
The result of Eq. (\ref{imveff}) follows directly from the calculations in Refs.~\cite{Labun:2012jf,Rafelski:2022bsv}. A full account will be presented in a review in preparation. 

To extend to any value of $G$, choose $G^\prime$ using $G=G^\prime+4k$  for some value of $k$ so that the value $G^\prime\to G$ is in the allowed domain. Note that for $G\to 0$ the well-known result for spin-0 mesons is found, and for $G\to 2$, one recovers the standard result for spin-$\tfrac{1}{2}$ particles (i.e. fermions such as quarks).(conventional result when setting fields to be all EM sourced). In this approach, the final
distribution of created quark pairs is grand-canonical. Even though we look at micro-canonical production processes in each heavy-ion
collision, the outcome is grand-canonical when the event ensemble is
considered. 

While all input elements to study isospin asymmetry are now in hand, considerable effort will be needed to obtain reliable model results for the relatively large expected isospin asymmetry, which can show a collision-energy and impact-parameter dependence in relativistic heavy-ion collisions.

Considering that SU(2)-flavor is a global (not gauge) approximate symmetry of QCD, there is no local constraint enforcing isospin neutrality. Consequently, microscopic processes can generate significant local isospin asymmetries, allowing the isospin content of subsystems to become large.
Since the hadronization process conserves isospin, this will impose additional constraints on small reaction systems. This is an additional difficulty for small collision systems.

The reader should keep in mind that the mechanism proposed for lepton-induced reactions ($e^+,e^-$) leading to isospin violation is very different from that proposed for relativistic heavy-ion collisions. However, in both cases the cause is the isospin-breaking electromagnetic interaction.

%% file: sections/appM.tex
We consider the possibility~\cite{Mrowczynski_ISOBREAK25} that the difference between charged and neutral kaon production originates from quark--antiquark pair creation in QCD, influenced by explicit $ud$-symmetry breaking due to the mass difference between up and down quarks. In particular, the interaction between the produced quark and antiquark, together with the $u$--$d$ mass difference, may lead to the observed difference.

When production of electron--positron pairs is considered, the produced electron and positron are usually treated as free particles. In his doctoral thesis, Andrei Sakharov studied the effect of the final-state Coulomb interaction of the electron and positron~\cite{Sakharov:1948plh}. The effect is significant and non-perturbative in the case of small relative momentum, when the electron and positron do not quickly fly apart, but remain within a distance of the order of the Bohr radius of the pair for a longer time. 

Sakharov concluded that, taking into account the interaction between the electron and positron, one obtains the product of the pair-production cross section, which neglects final-state interactions, with the modulus squared of the electron--positron pair wave function in the Coulomb scattering state. Since the pair production is localized in the region of the order of the inverse electron mass, which is much smaller than the Bohr radius, it is enough to take this wave function at a zero distance. 

The same result, although obtained in a slightly different context, is also called the Sommerfeld amplification factor, the Gamow factor, or the Fermi correction. The factor 
equals~\cite{Landau:1991wop}
\begin{equation}
\label{eq:Gamow}
\Gamma(p) \equiv |\psi_p ({\bf r}=0)|^2 = \pm \frac{2 \pi} {a_B p} \,
\frac{1}{{\exp}\big(\pm \frac{2 \pi}{a_B p}\big) - 1}\, ,
\end{equation}
where the sign plus (minus) is for the repulsive (attractive) Coulomb interaction, $p$ is the relative momentum of the particles, and $a_B$ is the Bohr radius of the pair. For $e^+e^-$ pairs, we consider only the attractive interaction. 

The quark--antiquark strong interaction is of the Coulomb form if the relative distance is much smaller than the inverse QCD scale parameter $\Lambda_{\rm QCD}^{-1} \approx 1$~fm. The potential, which is attractive when the $q\overline{q}$ pair is in the color singlet state and repulsive in the octet one, reads~\cite{Kniehl:2004rk}
\begin{equation}
\label{eq:Coulomb-pot}
V_1(r)  = - \frac{4}{3} \,\frac{\alpha_{\rm s}}{r}\, , ~~~~~~~~~~
V_8(r) =   \frac{1}{6} \, \frac{\alpha_{\rm s}}{r}\, ,
\end{equation}
where $\alpha_{\rm s}$ is the strong coupling constant while $4/3$ and $1/6$ are the SU(3) color factors, which for a generic ${\rm SU}(N )$ gauge group  are $C_F$ and $C_A/2 - C_F$ with the Casimir invariants $C_F = (N^2 -1)/(2N)$ and $C_A = N$.

\begin{figure}
\centering
\includegraphics[width=8.7cm]{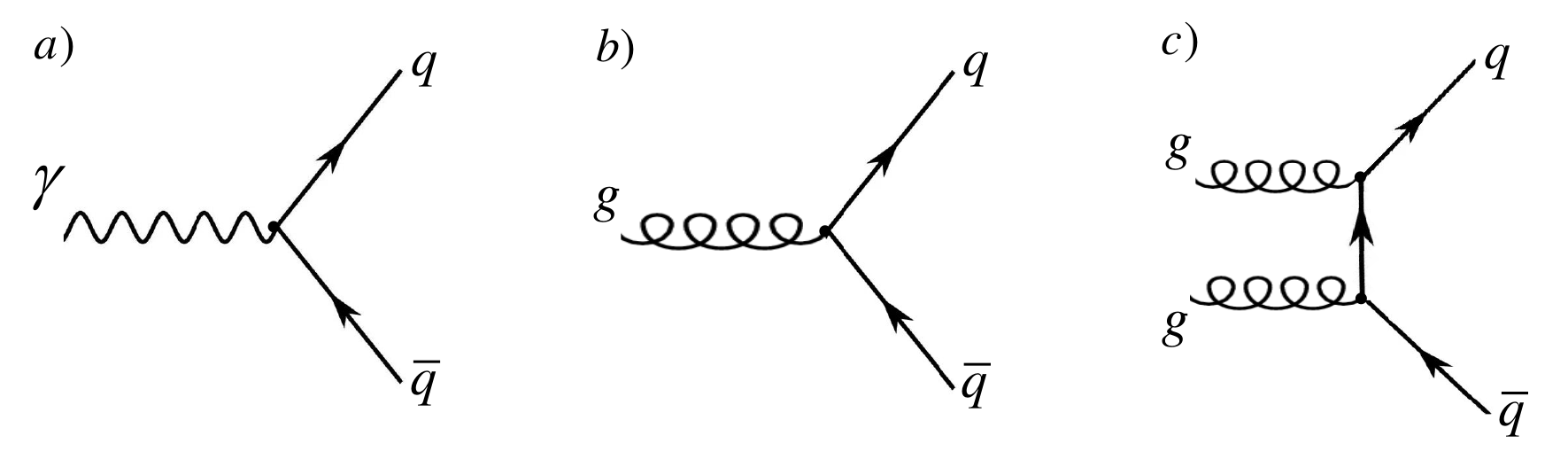}
\vspace{-3mm}
\caption{Three production processes of $q\overline{q}$ pair.}
\label{fig:diagrams}
\end{figure}

The color state of a $q\overline{q}$ pair depends on the production process. In the case of the electromagnetic production shown in Fig.~\ref{fig:diagrams}~(\textit{left}), which is relevant for \ee colliders, a virtual photon decays into the $q\overline{q}$ pair in the singlet state. The analogous one-gluon QCD process depicted in Fig.~\ref{fig:diagrams}~(\textit{middle}) produces the octet state.  In the case of the two-gluon process from Fig.~\ref{fig:diagrams} (\textit{right}), the $q\overline{q}$ pair can be in both singlet and octet states. Since the one-gluon process is of the lowest order in $\alpha_{\rm s}$, the octet channel is expected to dominate in the perturbative QCD. This observation is the basis of the Color Evaporation Model of quarkonium production~\cite{Gavai:1994in}. The model assumes that a $q\overline{q}$ pair is produced in the octet state via the one-gluon decay. Subsequently, it becomes color neutral by emitting a soft gluon. 

If the $q\overline{q}$ interaction is of the Coulomb form, the Gamow factor (Eq.~(\ref{eq:Gamow})) can be used to estimate the difference in production of $u\overline{u}$ and $d\overline{d}$ quark pairs due to their different masses and electromagnetic charges.

The inverse Bohr radii for the singlet and octet states, which enter the Gamow factor~(Eq.~(\ref{eq:Gamow})), are given by
\begin{equation}
\label{eq:a_B}
\frac{1}{a_B} =
\left\{
\begin{array}{ll}
\left(\dfrac{4}{3}\,\alpha_{\rm s} + q^2 \alpha_{\rm em}\right)\mu_R \,, & \text{singlet}\,, \\[2mm]
\left(\dfrac{1}{6}\,\alpha_{\rm s} - q^2 \alpha_{\rm em}\right)\mu_R \,, & \text{octet}\,,
\end{array}
\right.
\end{equation}
where $q$ is the quark electric charge in units of the fractional elementary charges ($q=2/3$ for $u$ quarks and $q=-1/3$ for $d$ quarks), $\alpha_{\rm em} = 1/137$ is the fine-structure constant, and $\mu_R$ denotes the reduced mass of the pair.
It equals $\mu = m_q/2$ where $m_q$ is the quark mass. Assuming that the current quark masses are relevant for the problem under consideration, we have 
$m_u = 2.16 \pm 0.07$~\MeV and $m_d = 4.70 \pm 0.07$~\MeV~\cite{ParticleDataGroup:2024cfk}. 
The sign plus in Eq.~(\ref{eq:a_B}) is for the attractive strong interaction and minus for the repulsive interaction. We further use $\alpha_{\rm s} = 0.3$, which is a typical value in a strong interaction phenomenology. Since $\alpha_{\rm s}/6 > q^2 \alpha_{\rm em}$, the Bohr radius is positive. 

\begin{figure}
\centering
\vspace{3mm}
\includegraphics[width=8.5cm]{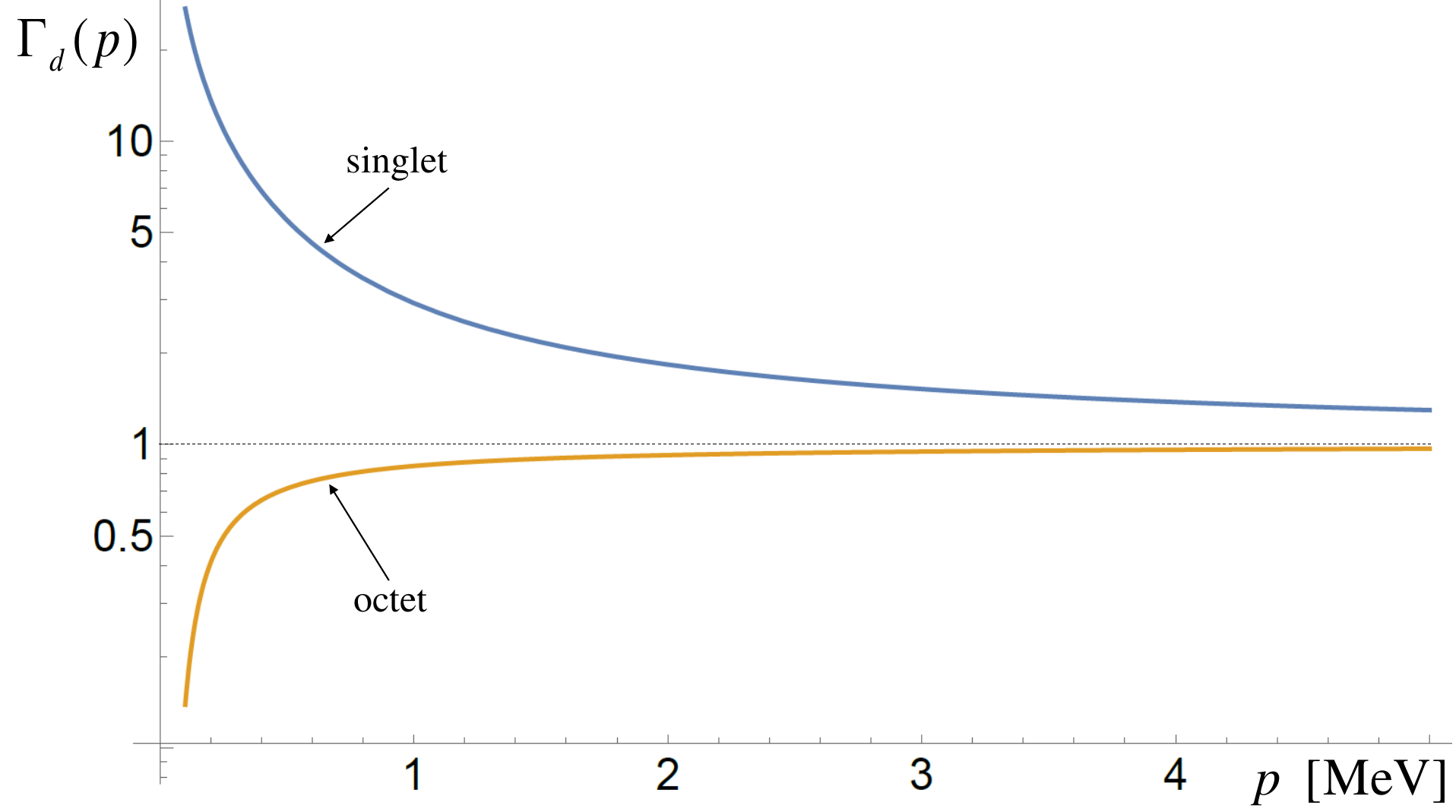} 
\includegraphics[width=8.5cm]{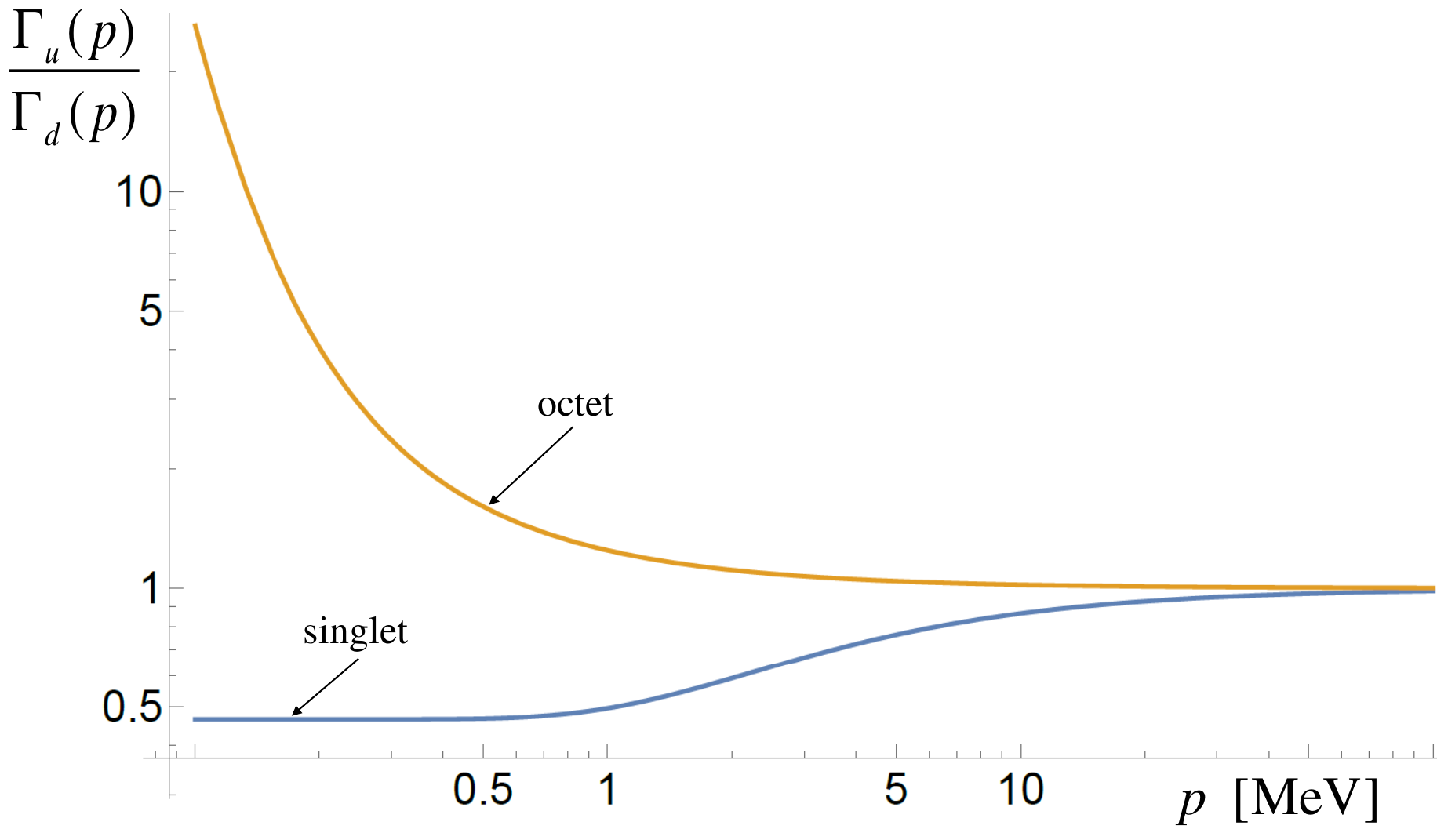}
\caption{\textit{Left}: The Gamow factors of  $d\overline{d}$ pair in the singlet and octet state as a function of relative momentum.
\textit{Right}: The ratios of Gamow factors of $u\overline{u}$ and  $d\overline{d}$ quark pairs in the singlet and octet states as a function of relative momentum.
The plots are taken from Ref.~\cite{Mrowczynski_ISOBREAK25}.}
\label{fig:singlet-vs-octet}
\end{figure}

In Fig.~\ref{fig:singlet-vs-octet} (\textit{left}), the Gamow factors of the $d\overline{d}$ pair in the singlet and octet states are shown. As one observes, the factors significantly differ from unity at small momenta for which the $d\overline{d}$ production is enhanced when the pair is in the singlet state and suppressed in the octet one. 
The ratios of Gamow factors of $u\overline{u}$ and $d\overline{d}$ quark pairs as a function of relative momentum are shown in 
Fig.~\ref{fig:singlet-vs-octet} (\textit{right}).
The pairs are either in the singlet or octet state. One sees that the production of up quarks is suppressed when compared to the down quarks if the quark--antiquark pairs are in the singlet state. This happens because the Bohr radius of $u\overline{u}$ is approximately two times bigger than that of $d\overline{d}$ because of the mass difference. The effect of different electromagnetic charges is completely negligible. At the vanishing relative momentum the ratio  $\Gamma_u(p)/\Gamma_d(p)$ in the singlet channel is minimal and it equals $\Gamma_u(0)/\Gamma_d(0) = a_B^d/a_B^u =  m_u/ m_d \approx 0.5$, as seen in Fig.~\ref{fig:singlet-vs-octet} (\textit{right}). In the case of the octet state, we observe the desired effect of enhanced production of $u$ quarks compared to $d$ quarks, which is the main result of this note.

Let us summarize the above considerations. A significant enhancement of $u\overline{u}$ production relative to $d\overline{d}$ may arise if 
\begin{itemize} 
\item quark pairs are produced in a color octet state,
\item their relative momentum is small. 
\end{itemize}

As already noted, the first condition arises naturally within perturbative QCD. The second is satisfied if quark--antiquark pairs are produced near threshold by soft gluons. In this sense, neither assumption is implausible. The main limitation of the analysis, however, is that it relies on arguments valid in the perturbative regime of QCD or for heavy quarks, but not in general. For light $u$ and $d$ quarks, the Bohr radius is not smaller than the inverse QCD scale, $\Lambda_{\rm QCD}^{-1}$, which calls into question the applicability of the Coulomb potentials~(\ref{eq:Coulomb-pot}). Moreover, Sakharov's key assumption -- that quark-pair production factorizes from subsequent interactions -- is itself uncertain. Consequently, the overall picture should be regarded with caution.

%% file: references.bib
@article{Brylinski:2023nrb,
    author = "Brylinski, Wojciech and Gazdzicki, Marek and Giacosa, Francesco and Gorenstein, Mark and Poberezhnyuk, Roman and Samanta, Subhasis",
    title = "{Evidence of isospin-symmetry violation in high-energy collisions of atomic nuclei: theoretical and phenomenological considerations}",
    eprint = "2312.07176",
    archivePrefix = "arXiv",
    primaryClass = "nucl-th",
    doi = "10.5506/APhysPolB.57.3-A5",
    journal = "Acta Phys. Polon. B",
    volume = "57",
    number = "3",
    pages = "3--A5",
    year = "2026"
}

@article{Letessier:2005qe,
    author = "Letessier, Jean and Rafelski, Johann",
    title = "{Hadron production and phase changes in relativistic heavy-ion collisions}",
    eprint = "nucl-th/0504028",
    archivePrefix = "arXiv",
    reportNumber = "CERN-PH-TH-2005-060",
    doi = "10.1140/epja/i2007-10546-7",
    journal = "Eur. Phys. J. A",
    volume = "35",
    pages = "221--242",
    year = "2008"
}

@article{Anselm:1991pi,
    author = "Anselm, A. A. and Ryskin, M. G.",
    title = "{Production of classical pion field in heavy ion high energy collisions}",
    doi = "10.1016/0370-2693(91)91073-5",
    journal = "Phys. Lett. B",
    volume = "266",
    pages = "482--484",
    year = "1991"
}

@article{Blaizot:1992at,
    author = "Blaizot, Jean-Paul and Krzywicki, André",
    title = "{Soft-pion emission in high-energy heavy-ion collisions}",
    reportNumber = "LPTHE-ORSAY-92-11",
    doi = "10.1103/PhysRevD.46.246",
    journal = "Phys. Rev. D",
    volume = "46",
    pages = "246--251",
    year = "1992"
}

@article{Bjorken:1993wj,
    author = "Bjorken, J. D. and Kowalski, K. L. and Taylor, C. C.",
    title = "{Observing disoriented chiral condensates}",
    booktitle = "{Workshop on Physics at Current Accelerators and the Supercollider}",
    eprint = "hep-ph/9309235",
    archivePrefix = "arXiv",
    reportNumber = "SLAC-PUB-6413",
    year = "1993"
}

@article{Rajagopal:1993ah,
    author = "Rajagopal, Krishna and Wilczek, Frank",
    title = "{Emergence of coherent long wavelength oscillations after a quench: application to QCD}",
    eprint = "hep-ph/9303281",
    archivePrefix = "arXiv",
    reportNumber = "PUPT-1389, IASSNS-HEP-93-16",
    doi = "10.1016/0550-3213(93)90591-C",
    journal = "Nucl. Phys. B",
    volume = "404",
    pages = "577--589",
    year = "1993"
}

@article{Schaffner-Bielich:1998mra,
    author = "Schaffner-Bielich, Jurgen and Randrup, Jorgen",
    title = "{Disoriented chiral condensate dynamics with the SU(3) linear sigma model}",
    eprint = "nucl-th/9812032",
    archivePrefix = "arXiv",
    reportNumber = "LBNL-42018, LBL-42018",
    doi = "10.1103/PhysRevC.59.3329",
    journal = "Phys. Rev. C",
    volume = "59",
    pages = "3329--3342",
    year = "1999"
}

@article{Gavin:2001uk,
    author = "Gavin, Sean and Kapusta, Joseph I.",
    title = "{Kaon and pion fluctuations from small disoriented chiral condensates}",
    eprint = "nucl-th/0112083",
    archivePrefix = "arXiv",
    reportNumber = "NUC-MINN-01-20-T",
    doi = "10.1103/PhysRevC.65.054910",
    journal = "Phys. Rev. C",
    volume = "65",
    pages = "054910",
    year = "2002"
}

@article{Kapusta:2023xrw,
    author = "Kapusta, Joseph I. and Pratt, Scott and Singh, Mayank",
    title = "{Disoriented isospin condensates may be the source of anomalous kaon correlations measured in Pb-Pb collisions at $\sqrt{s_{NN}}$ $=$ 2.76 TeV}",
    eprint = "2306.13280",
    archivePrefix = "arXiv",
    primaryClass = "hep-ph",
    doi = "10.1103/PhysRevC.109.L031902",
    journal = "Phys. Rev. C",
    volume = "109",
    number = "3",
    pages = "L031902",
    year = "2024"
}

@article{Rapp:1999ej,
    author = "Rapp, R. and Wambach, J.",
    title = "{Chiral symmetry restoration and dileptons in relativistic heavy-ion collisions}",
    eprint = "hep-ph/9909229",
    archivePrefix = "arXiv",
    reportNumber = "SUNY-NTG-99-29",
    doi = "10.1007/0-306-47101-9_1",
    journal = "Adv. Nucl. Phys.",
    volume = "25",
    pages = "1",
    year = "2000"
}

@article{Ko:1997kb,
    author = "Ko, Che Ming and Koch, Volker and Li, Guo-Qiang",
    title = "{Properties of hadrons in the nuclear medium}",
    eprint = "nucl-th/9702016",
    archivePrefix = "arXiv",
    reportNumber = "LBL-39866, LBNL-39866",
    doi = "10.1146/annurev.nucl.47.1.505",
    journal = "Ann. Rev. Nucl. Part. Sci.",
    volume = "47",
    pages = "505--539",
    year = "1997"
}

@article{Fuchs:2005zg,
    author = "Fuchs, Christian",
    title = "{Kaon production in heavy ion reactions at intermediate energies}",
    eprint = "nucl-th/0507017",
    archivePrefix = "arXiv",
    doi = "10.1016/j.ppnp.2005.07.004",
    journal = "Prog. Part. Nucl. Phys.",
    volume = "56",
    pages = "1--103",
    year = "2006"
}

@inproceedings{SQM24,
    author = "Singh, M. and others",
    booktitle = "{21st International Conference on Strangeness in Quark Matter (SQM 2024)}",
    title = "{Disoriented Isospin Condensates as
source of anomalous kaon correlations at LHC}",
    url = "https://indico.in2p3.fr/event/29792/contributions/137151",
    year = "2024"
}

@article{Gao:2024nkz,
    author = "Gao, Jun and Liu, ChongYang and Shen, XiaoMin and Xing, Hongxi and Zhao, Yuxiang",
    title = "{Simultaneous Determination of Fragmentation Functions and Test on Momentum Sum Rule}",
    eprint = "2401.02781",
    archivePrefix = "arXiv",
    primaryClass = "hep-ph",
    doi = "10.1103/PhysRevLett.132.261903",
    journal = "Phys. Rev. Lett.",
    volume = "132",
    number = "26",
    pages = "261903",
    year = "2024"
}

@article{Gao:2024dbv,
    author = "Gao, Jun and Liu, ChongYang and Shen, XiaoMin and Xing, Hongxi and Zhao, Yuxiang",
    title = "{Global analysis of fragmentation functions to charged hadrons with high-precision data from the LHC}",
    eprint = "2407.04422",
    archivePrefix = "arXiv",
    primaryClass = "hep-ph",
    doi = "10.1103/PhysRevD.110.114019",
    journal = "Phys. Rev. D",
    volume = "110",
    number = "11",
    pages = "114019",
    year = "2024"
}

@article{BESIII:2025mbc,
    author = "Ablikim, Medina and others",
    collaboration = "BESIII",
    title = "{Single Inclusive $\pi^\pm$ and $K^\pm$ Production in $e^+e^-$ Annihilation at Center-of-Mass Energies from 2.000 to 3.671 GeV}",
    eprint = "2502.16084",
    archivePrefix = "arXiv",
    primaryClass = "hep-ex",
    doi = "10.1103/5kxw-3mbw",
    journal = "Phys. Rev. Lett.",
    volume = "135",
    pages = "151901",
    year = "2025"
}

@article{Bhatt:2024prq,
    author = "Bhatt, H. and others",
    title = "{Flavor dependence of charged pion fragmentation functions}",
    eprint = "2408.16640",
    archivePrefix = "arXiv",
    primaryClass = "nucl-ex",
    doi = "10.1016/j.physletb.2025.139485",
    journal = "Phys. Lett. B",
    volume = "865",
    pages = "139485",
    year = "2025"
}

@article{NA61SHINE:2023azp,
    author = "Adhikary, H. and others",
    collaboration = "NA61/SHINE",
    title = "{Evidence of isospin-symmetry violation in high-energy collisions of atomic nuclei}",
    eprint = "2312.06572",
    archivePrefix = "arXiv",
    primaryClass = "nucl-ex",
    reportNumber = "CERN-EP-2023-283",
    doi = "10.1038/s41467-025-57234-6",
    journal = "Nature Commun.",
    volume = "16",
    number = "1",
    pages = "2849",
    year = "2025"
}

@unpublished{Kalisky:thesis,
    author = "Mat\'u\v{s} Kalisk\'y",
    title = "{{Reconstruction of charged kaons in the three pion decay channel in Pb+Au 158 AGeV collisions by the CERES experiment}}",
    note		="PhD thesis, Technical University of Darmstadt",
    year = "2007",
    url="https://cds.cern.ch/record/1497739"
}

@article{STAR:2019bjj,
    author = "Adam, Jaroslav and others",
    collaboration = "STAR",
    title = "{Strange hadron production in Au+Au collisions at $\sqrt{s_{_{\mathrm{NN}}}}$ = 7.7, 11.5, 19.6, 27, and 39 GeV}",
    eprint = "1906.03732",
    archivePrefix = "arXiv",
    primaryClass = "nucl-ex",
    doi = "10.1103/PhysRevC.102.034909",
    journal = "Phys. Rev. C",
    volume = "102",
    number = "3",
    pages = "034909",
    year = "2020"
}

@article{STAR:2017sal,
    author = "Adamczyk, L. and others",
    collaboration = "STAR",
    title = "{Bulk properties of the medium produced in relativistic heavy-ion collisions from the beam energy scan program}",
    eprint = "1701.07065",
    archivePrefix = "arXiv",
    primaryClass = "nucl-ex",
    doi = "10.1103/PhysRevC.96.044904",
    journal = "Phys. Rev. C",
    volume = "96",
    number = "4",
    pages = "044904",
    year = "2017"
}

@article{STAR:2002hpr,
    author = "Adler, C. and others",
    collaboration = "STAR",
    title = "{Kaon production and kaon to pion ratio in Au+Au collisions at $\sqrt{s_{NN}}$ $=$ 130 GeV}",
    eprint = "nucl-ex/0206008",
    archivePrefix = "arXiv",
    doi = "10.1016/j.physletb.2004.06.044",
    journal = "Phys. Lett. B",
    volume = "595",
    pages = "143--150",
    year = "2004"
}

@article{STAR:2010yyv,
    author = "Aggarwal, M. M. and others",
    collaboration = "STAR",
    title = "{Strange and multistrange particle production in Au+Au collisions at $\sqrt{s_{NN}}$ $=$ 62.4 GeV}",
    eprint = "1010.0142",
    archivePrefix = "arXiv",
    primaryClass = "nucl-ex",
    doi = "10.1103/PhysRevC.83.024901",
    journal = "Phys. Rev. C",
    volume = "83",
    pages = "024901",
    year = "2011"
}

@article{STAR:2011fbd,
    author = "Agakishiev, G. and others",
    collaboration = "STAR",
    title = "{Strangeness Enhancement in Cu+Cu and Au+Au Collisions at $\sqrt{s_{NN}} = 200$ GeV}",
    eprint = "1107.2955",
    archivePrefix = "arXiv",
    primaryClass = "nucl-ex",
    doi = "10.1103/PhysRevLett.108.072301",
    journal = "Phys. Rev. Lett.",
    volume = "108",
    pages = "072301",
    year = "2012"
}

@article{STAR:2008med,
    author = "Abelev, B. I. and others",
    collaboration = "STAR",
    title = "{Systematic measurements of identified particle spectra in $pp$, $d$+Au and Au+Au collisions at the STAR detector}",
    eprint = "0808.2041",
    archivePrefix = "arXiv",
    primaryClass = "nucl-ex",
    doi = "10.1103/PhysRevC.79.034909",
    journal = "Phys. Rev. C",
    volume = "79",
    pages = "034909",
    year = "2009"
}

@article{ALICE:2013cdo,
    author = "Abelev, Betty and others",
    collaboration = "ALICE",
    title = "{$K^0_S$ and $\Lambda$ Production in Pb-Pb Collisions at $\sqrt{s_{NN}}$ = 2.76 TeV}",
    eprint = "1307.5530",
    archivePrefix = "arXiv",
    primaryClass = "nucl-ex",
    reportNumber = "CERN-PH-EP-2013-132",
    doi = "10.1103/PhysRevLett.111.222301",
    journal = "Phys. Rev. Lett.",
    volume = "111",
    pages = "222301",
    year = "2013"
}

@article{ALICE:2013mez,
    author = "Abelev, Betty and others",
    collaboration = "ALICE",
    title = "{Centrality dependence of $\pi$, $K$, and $p$ production in Pb-Pb collisions at $\sqrt{s_{NN}}$ = 2.76 TeV}",
    eprint = "1303.0737",
    archivePrefix = "arXiv",
    primaryClass = "hep-ex",
    reportNumber = "CERN-PH-EP-2013-019",
    doi = "10.1103/PhysRevC.88.044910",
    journal = "Phys. Rev. C",
    volume = "88",
    pages = "044910",
    year = "2013"
}

@article{NA35:1994veo,
    author = "Alber, T. and others",
    collaboration = "NA35",
    title = "{Strange particle production in nuclear collisions at 200 GeV per nucleon}",
    doi = "10.1007/BF01557391",
    journal = "Z. Phys. C",
    volume = "64",
    pages = "195--207",
    year = "1994"
}

@article{NA35:1992njd,
    author = "Baechler, J. and others",
    collaboration = "NA35",
    title = "{Production of charged kaons in proton-nucleus and nucleus-nucleus collisions at 200 GeV/nucleon}",
    reportNumber = "MPI-PHE-92-21",
    doi = "10.1007/BF01557694",
    journal = "Z. Phys. C",
    volume = "58",
    pages = "367--374",
    year = "1993"
}

@article{HADES:2009lnd,
    author = "Agakishiev, G. and others",
    collaboration = "HADES",
    title = "{$\phi$ decay: A relevant source for K$^-$ production at energies available at the GSI Schwerionen-Synchrotron (SIS)?}",
    eprint = "0902.3487",
    archivePrefix = "arXiv",
    primaryClass = "nucl-ex",
    doi = "10.1103/PhysRevC.80.025209",
    journal = "Phys. Rev. C",
    volume = "80",
    pages = "025209",
    year = "2009"
}

@article{Agakishiev:2010zw,
    author = "Agakishiev, G. and others",
    title = "{In-medium effects on $K^0$ mesons in relativistic heavy-ion collisions}",
    eprint = "1004.3881",
    archivePrefix = "arXiv",
    primaryClass = "nucl-ex",
    doi = "10.1103/PhysRevC.82.044907",
    journal = "Phys. Rev. C",
    volume = "82",
    pages = "044907",
    year = "2010"
}

@article{Petran:2013lja,
    author = "Petr\'a\v{n}, Michal and Letessier, Jean and Petr\'a\v{c}ek, Vojt\v{e}ch and Rafelski, Johann",
    title = "{Hadron production and quark-gluon plasma hadronization in Pb-Pb collisions at $\sqrt{s_{NN}}=2.76$ TeV}",
    eprint = "1303.2098",
    archivePrefix = "arXiv",
    primaryClass = "hep-ph",
    reportNumber = "CERN-PH-TH-2012-262",
    doi = "10.1103/PhysRevC.88.034907",
    journal = "Phys. Rev. C",
    volume = "88",
    number = "3",
    pages = "034907",
    year = "2013"
}

@article{Rafelski:2014cqa,
    author = "Rafelski, Johann and Petr\'a\v{n}, Michal",
    title = "{Universal QGP Hadronization Conditions at RHIC and LHC}",
    eprint = "1406.1871",
    archivePrefix = "arXiv",
    primaryClass = "nucl-th",
    doi = "10.1051/epjconf/20147806004",
    journal = "EPJ Web Conf.",
    volume = "78",
    pages = "06004",
    year = "2014"
}

@article{Rafelski:2015hta,
    author = "Rafelski, Jan and Petran, Michal",
    title = "{QCD Phase Transition Studied by Means of Hadron Production}",
    eprint = "2212.13302",
    archivePrefix = "arXiv",
    primaryClass = "hep-ph",
    doi = "10.1134/S1063779615050238",
    journal = "Phys. Part. Nucl.",
    volume = "46",
    number = "5",
    pages = "748--755",
    year = "2015"
}

@article{Bleicher:1999xi,
    author = "Bleicher, M. and others",
    title = "{Relativistic hadron-hadron collisions in the ultrarelativistic quantum molecular dynamics model}",
    eprint = "hep-ph/9909407",
    archivePrefix = "arXiv",
    doi = "10.1088/0954-3899/25/9/308",
    journal = "J. Phys. G",
    volume = "25",
    pages = "1859--1896",
    year = "1999"
}

@article{Bleicher:2022kcu,
    author = "Bleicher, Marcus and Bratkovskaya, Elena",
    title = "{Modelling relativistic heavy-ion collisions with dynamical transport approaches}",
    doi = "10.1016/j.ppnp.2021.103920",
    journal = "Prog. Part. Nucl. Phys.",
    volume = "122",
    pages = "103920",
    year = "2022"
}

@article{Pisarski:1983ms,
    author = "Pisarski, Robert D. and Wilczek, Frank",
    title = "{Remarks on the chiral phase transition in chromodynamics}",
    reportNumber = "NSF-ITP-83-152",
    doi = "10.1103/PhysRevD.29.338",
    journal = "Phys. Rev. D",
    volume = "29",
    pages = "338--341",
    year = "1984"
}

@article{Stepaniak:2023pvo,
    author = "Stepaniak, Joanna and Pszczel, Damian",
    title = "{On the relation between $K^{0}_{s}$ and charged kaon yields in proton{\textendash}proton collisions}",
    eprint = "2305.03872",
    archivePrefix = "arXiv",
    primaryClass = "hep-ph",
    doi = "10.1140/epjc/s10052-023-12108-3",
    journal = "Eur. Phys. J. C",
    volume = "83",
    number = "10",
    pages = "928",
    year = "2023"
}

@techreport{Mackowiak-Pawlowska:2867952,
      author        = "Adhikary, H and others",
      title         = "{Addendum to the NA61/SHINE Proposal: Request for light ions beams in Run 4}",
      institution   = "CERN",
      collaboration = "NA61/SHINE",
      address       = "Geneva",
      number        = "CERN-SPSC-2023-022, SPSC-P-330-ADD-14",
      year          = "2023",
      reportNumber  = "CERN-SPSC-2023-022",
      url           = "https://cds.cern.ch/record/2867952",
}

@techreport{Kowalski:2907307,
author        = "NA61/SHINE",
      collaboration = "NA61/SHINE",
      title         = "{Memorandum requesting use of the allocated test beam for
                       data-taking on $\pi^{+}$+ C and $\pi^{-}$+ C interactions
                       at 158 GeV/$c$}",
      institution   = "CERN",
      reportNumber  = "CERN-SPSC-2024-022",
      number        = "CERN-SPSC-2024-022, SPSC-M-797",
      address       = "Geneva",
      year          = "2024",
      url           = "https://cds.cern.ch/record/2907307",
}

@article{ALICE:2021fpb,
    author = "Acharya, Shreyasi and others",
    collaboration = "ALICE",
    title = "{Neutral to charged kaon yield fluctuations in Pb \textendash{} Pb collisions at \snn=2.76 TeV}",
    eprint = "2112.09482",
    archivePrefix = "arXiv",
    primaryClass = "nucl-ex",
    reportNumber = "CERN-EP-2021-261",
    doi = "10.1016/j.physletb.2022.137242",
    journal = "Phys. Lett. B",
    volume = "832",
    pages = "137242",
    year = "2022"
}

@article{Kapusta:2022ovq,
    author = "Kapusta, Joseph I. and Pratt, Scott and Singh, Mayank",
    title = "{Confronting anomalous kaon correlations measured in Pb-Pb collisions at $\sqrt{s_{NN}}=2.76$ TeV}",
    eprint = "2210.03257",
    archivePrefix = "arXiv",
    primaryClass = "hep-ph",
    doi = "10.1103/PhysRevC.107.014913",
    journal = "Phys. Rev. C",
    volume = "107",
    number = "1",
    pages = "014913",
    year = "2023"
}

@article{Pennington:2005be,
    author = "Pennington, M. R.",
    title = "{Swimming with Quarks}",
    eprint = "hep-ph/0504262",
    archivePrefix = "arXiv",
    doi = "10.1088/1742-6596/18/1/001",
    journal = "J. Phys. Conf. Ser.",
    volume = "18",
    pages = "1--73",
    year = "2005"
}

@article{Alarcon:2011zs,
    author = "Alarc\'on, J. M. and Martin Camalich, J. and Oller, J. A.",
    title = "{Chiral representation of the $\pi N$ scattering amplitude and the pion-nucleon sigma term}",
    eprint = "1110.3797",
    archivePrefix = "arXiv",
    primaryClass = "hep-ph",
    doi = "10.1103/PhysRevD.85.051503",
    journal = "Phys. Rev. D",
    volume = "85",
    pages = "051503(R)",
    year = "2012"
}

@article{ParticleDataGroup:2024cfk,
    author = "Navas, S. and others",
    collaboration = "Particle Data Group",
    title = "{Review of Particle Physics}",
    doi = "10.1103/PhysRevD.110.030001",
    journal = "Phys. Rev. D",
    volume = "110",
    number = "3",
    pages = "030001",
    year = "2024"
}

@article{Giacosa:2023fdz,
    author = "Giacosa, Francesco and Jafarzade, Shahriyar and Pisarski, Robert D.",
    title = "{Anomalous interactions between mesons with nonzero spin and glueballs}",
    eprint = "2309.00086",
    archivePrefix = "arXiv",
    primaryClass = "hep-ph",
    doi = "10.1103/PhysRevD.109.L071502",
    journal = "Phys. Rev. D",
    volume = "109",
    number = "7",
    pages = "L071502",
    year = "2024"
}

@article{Bonesini:2001iz,
    author = "Bonesini, M. and Marchionni, A. and Pietropaolo, F. and Tabarelli de Fatis, T.",
    title = "{On particle production for high energy neutrino beams}",
    eprint = "hep-ph/0101163",
    archivePrefix = "arXiv",
    reportNumber = "CERN-SL-2001-005-EA",
    doi = "10.1007/s100520100656",
    journal = "Eur. Phys. J. C",
    volume = "20",
    pages = "13--27",
    year = "2001"
}

@article{Giacosa:2025ynn,
    author = "Giacosa, Francesco and Rohrmoser, Martin",
    title = "{Isospin kaon anomaly and its consequences}",
    eprint = "2504.02113",
    archivePrefix = "arXiv",
    primaryClass = "nucl-th",
    doi = "10.1140/epjc/s10052-025-14798-3",
    journal = "Eur. Phys. J. C",
    volume = "85",
    number = "9",
    pages = "1058",
    year = "2025"
}

@article{Giacosa:2024bup,
    author = "Giacosa, Francesco",
    title = "{Isospin-symmetry breaking in kaon production: the role of charge-symmetry}",
    eprint = "2409.14763",
    archivePrefix = "arXiv",
    primaryClass = "nucl-th",
    doi = "10.1051/epjconf/202531603002",
    journal = "EPJ Web Conf.",
    volume = "316",
    pages = "03002",
    year = "2025"
}

@article{Reichert:2025znn,
    author = "Reichert, Tom and Steinheimer, Jan and Bleicher, Marcus",
    title = "{Explanation of the observed violation of isospin symmetry in relativistic nucleus-nucleus reactions}",
    eprint = "2503.10493",
    archivePrefix = "arXiv",
    primaryClass = "nucl-th",
    doi = "10.1016/j.physletb.2026.140170",
    journal = "Phys. Lett. B",
    volume = "873",
    pages = "140170",
    year = "2026"
}

@article{Kovacs:2024cdb,
    author = {Kov\'acs, P\'eter and Wolf, Gy\"orgy and Weickgenannt, Nora and Rischke, Dirk H.},
    title = "{Phenomenology of isospin-symmetry breaking with vector mesons}",
    eprint = "2401.04527",
    archivePrefix = "arXiv",
    primaryClass = "hep-ph",
    doi = "10.1103/PhysRevD.109.096007",
    journal = "Phys. Rev. D",
    volume = "109",
    number = "9",
    pages = "096007",
    year = "2024"
}

@article{Kharzeev:1996sq,
    author = "Kharzeev, D.",
    title = "{Can gluons trace baryon number?}",
    eprint = "nucl-th/9602027",
    archivePrefix = "arXiv",
    reportNumber = "CERN-TH-95-343, BI-TP-95-42",
    doi = "10.1016/0370-2693(96)00435-2",
    journal = "Phys. Lett. B",
    volume = "378",
    pages = "238--246",
    year = "1996"
}

@article{Pihan:2024lxw,
    author = {Pihan, Gregoire and Monnai, Akihiko and Schenke, Bj{\"o}rn and Shen, Chun},
    title = "{Unveiling Baryon Charge Carriers through Charge Stopping in Isobar Collisions}",
    eprint = "2405.19439",
    archivePrefix = "arXiv",
    primaryClass = "nucl-th",
    doi = "10.1103/PhysRevLett.133.182301",
    journal = "Phys. Rev. Lett.",
    volume = "133",
    number = "18",
    pages = "182301",
    year = "2024"
}

@article{SeaQuest:2021zxb,
    author = "Dove, J. and others",
    collaboration = "SeaQuest",
    title = "{The asymmetry of antimatter in the proton}",
    eprint = "2103.04024",
    archivePrefix = "arXiv",
    primaryClass = "hep-ph",
    reportNumber = "FERMILAB-PUB-21-073-E",
    doi = "10.1038/s41586-022-04707-z",
    journal = "Nature",
    volume = "590",
    number = "7847",
    pages = "561--565",
    year = "2021",
    note = "[Erratum: Nature 604, E26 (2022)]"
}

@article{Geesaman:2018ixo,
    author = "Geesaman, D. F. and Reimer, P. E.",
    title = "{The sea of quarks and antiquarks in the nucleon}",
    eprint = "1812.10372",
    archivePrefix = "arXiv",
    primaryClass = "nucl-ex",
    doi = "10.1088/1361-6633/ab05a7",
    journal = "Rept. Prog. Phys.",
    volume = "82",
    number = "4",
    pages = "046301",
    year = "2019"
}

@article{Ikeno:2019mne,
    author = "Ikeno, Natsumi and Ono, Akira and Nara, Yasushi and Ohnishi, Akira",
    title = "{Effects of Pauli blocking on pion production in central collisions of neutron-rich nuclei}",
    eprint = "1910.09173",
    archivePrefix = "arXiv",
    primaryClass = "nucl-th",
    doi = "10.1103/PhysRevC.101.034607",
    journal = "Phys. Rev. C",
    volume = "101",
    number = "3",
    pages = "034607",
    year = "2020"
}

@article{Serpukhov-Brussels-AnnecyLAPP:1984xib,
    author = "Binon, F. and others",
    collaboration = "IHEP--IISN--LAP",
    title = "{The isospin-violating decay $\eta^\prime \to 3 \pi^0$}",
    reportNumber = "CERN-EP/84-27",
    doi = "10.1016/0370-2693(84)90932-8",
    journal = "Phys. Lett. B",
    volume = "140",
    pages = "264--268",
    year = "1984"
}

@article{Gross:1979ur,
    author = "Gross, David J. and Treiman, S. B. and Wilczek, Frank",
    title = "{Light-quark masses and isospin violation}",
    reportNumber = "Print-79-0123 (PRINCETON)",
    doi = "10.1103/PhysRevD.19.2188",
    journal = "Phys. Rev. D",
    volume = "19",
    pages = "2188",
    year = "1979"
}

@article{Balkova:2025dpf,
    author = "Balkova, Yuliia and {\v{S}}u{\v{s}}a, Tatjana",
    title = "{News on strangeness production from the NA61/SHINE experiment}",
    eprint = "2508.14998",
    archivePrefix = "arXiv",
    primaryClass = "nucl-ex",
    doi = "10.1051/epjconf/202636411003",
    journal = "EPJ Web Conf.",
    volume = "364",
    pages = "11003",
    year = "2026"
}

@article{NA61SHINE:2023epu,
    author = "Adhikary, H. and others",
    collaboration = "NA61/SHINE",
    title = "{Measurements of $\pi ^\pm$, $K^\pm $, $p$ and $\bar{p}$ spectra in $^{40}$Ar+$^{45}$Sc collisions at 13$A$ to 150$A$ GeV/$c$}",
    eprint = "2308.16683",
    archivePrefix = "arXiv",
    primaryClass = "nucl-ex",
    reportNumber = "CERN-EP-2023-179, FERMILAB-PUB-23-563-AD",
    doi = "10.1140/epjc/s10052-024-12602-2",
    journal = "Eur. Phys. J. C",
    volume = "84",
    number = "4",
    pages = "416",
    year = "2024"
}

@article{NA61SHINE:2017fne,
    author = "Aduszkiewicz, A. and others",
    collaboration = "NA61/SHINE",
    title = "{Measurements of $\pi ^\pm$ , K$^\pm$, p and ${\bar{\text {p}}}$ spectra in proton-proton interactions at 20, 31, 40, 80 and 158  GeV/$c$ with the NA61/SHINE spectrometer at the CERN SPS}",
    eprint = "1705.02467",
    archivePrefix = "arXiv",
    primaryClass = "nucl-ex",
    reportNumber = "CERN-EP-2017-066",
    doi = "10.1140/epjc/s10052-017-5260-4",
    journal = "Eur. Phys. J. C",
    volume = "77",
    number = "10",
    pages = "671",
    year = "2017"
}

@article{NA61SHINE:2020czq,
    author = "Acharya, A. and others",
    collaboration = "NA61/SHINE",
    title = "{Measurements of $\pi^\pm$, $K^\pm$, $p$ and $\bar{p}$ spectra in $^7$Be+$^9$Be collisions at beam momenta from 19$A$ to 150$A$ GeV/$c$ with the NA61/SHINE spectrometer at the CERN SPS}",
    eprint = "2010.01864",
    archivePrefix = "arXiv",
    primaryClass = "hep-ex",
    reportNumber = "CERN-EP-2020-187, CERN-EP-2020-187",
    doi = "10.1140/epjc/s10052-020-08733-x",
    journal = "Eur. Phys. J. C",
    volume = "81",
    number = "1",
    pages = "73",
    year = "2021",
    note = "[Erratum: Eur.Phys.J.C 83, 90 (2023)]"
}

@article{NA49:2002pzu,
    author = "Afanasiev, S. V. and others",
    collaboration = "NA49",
    title = "{Energy dependence of pion and kaon production in central Pb+Pb collisions}",
    eprint = "nucl-ex/0205002",
    archivePrefix = "arXiv",
    doi = "10.1103/PhysRevC.66.054902",
    journal = "Phys. Rev. C",
    volume = "66",
    pages = "054902",
    year = "2002"
}

@article{NA49:2007stj,
    author = "Alt, C. and others",
    collaboration = "NA49",
    title = "{Pion and kaon production in central Pb+Pb collisions at 20$A$ and 30$A$ GeV: Evidence for the onset of deconfinement}",
    eprint = "0710.0118",
    archivePrefix = "arXiv",
    primaryClass = "nucl-ex",
    doi = "10.1103/PhysRevC.77.024903",
    journal = "Phys. Rev. C",
    volume = "77",
    pages = "024903",
    year = "2008"
}

@article{Radomski:2008zz,
    author = "Radomski, S.",
    collaboration = "CERES",
    title = "{CERES measurement of strangeness production at top SPS energy}",
    doi = "10.1088/0954-3899/35/4/044003",
    journal = "J. Phys. G",
    volume = "35",
    pages = "044003",
    year = "2008"
}

@unpublished{Radomski:thesis,
    author = "Sylwester Radomski",
    title = "{{Neutral strange particle production at top SPS energy measured by the CERES experiment}}",
    note		="PhD thesis, Technical University of Darmstadt",
    year = "2006",
    url="http://cds.cern.ch/record/1497741"
}

@unpublished{Strabel:thesis,
    author = "Claudia Strabel",
    title = "{{Energieabhängigkeit der K$^0_\mathrm{S}$-Produktion in zentralen Pb+Pb Reaktionen}}",
    note		="Diploma thesis, Johann Wolfgang Goethe-Universität",
    year = "2006",
    url="https://edms.cern.ch/document/2958534/1"
}

@unpublished{Kutsche:thesis,
    author = "Ralf Kutsche",
    title = "{{Untersuchungen der In-Medium-Eigenschaften von $K^0_S$-Mesonen und $\ensuremath{\Lambda}$-Hyperonen an der Produktionsschwelle}}",
    note		="PhD thesis, Technische Universit{\"a}t Darmstadt",
    year = "1999",
    url = "https://tuprints.ulb.tu-darmstadt.de/entities/publication/1785882c-7b10-419f-8cc0-861d14893396"
}

@unpublished{Barnby:thesis,
    author = "Lee Stuart Barnby",
    title = "{{Measurements of $\Lambda$, $\overline{\Lambda}$ and
$K^0_S$ from Pb-Pb Collisions at 158 GeV per nucleon in a Large Acceptance Experiment}}",
    note		="PhD thesis, University of Birmingham",
    year = "1999",
    url = "https://edms.cern.ch/document/816018/1"
}

@unpublished{Mischke:thesis,
    author = "Andre Mischke",
    title = "{{$\Lambda$ und $\overline{\Lambda}$ Produktion in zentralen Blei-Blei-Kollisionen bei 40, 80 und 158 GeV pro Nukleon}}",
    note		="PhD thesis, Johann Wolfgang Goethe-Universität",
    year = "2002",
    url = "https://edms.cern.ch/document/3409580/1"
}

@unpublished{Book:thesis,
    author = "Julian Book",
    title = "{{Zentralitätsabhängigkeit der $K^0_S$-Produktion in relativistischen Schwerionenkollisionen}}",
    note		="Diploma thesis, Johann Wolfgang Goethe-Universität",
    year = "2009",
    url = "https://edms.cern.ch/document/3409578/1"
}

@inproceedings{Ercolessi_ISOBREAK25,
    author = "Ercolessi, Francesca",
    collaboration = "for ALICE",
    booktitle = "{Workshop on isospin symmetry violation: kaons and beyond -- ISO-BREAK 25}",
    title = "{Charged and neutral Kaon production in ALICE}",
    url = "https://indico.cern.ch/event/1557894/contributions/6698090",
    year = "2025",
}

@inproceedings{Turko_ISOBREAK25,
    author = "Turko, Ludwik",
%    collaboration = "for ALICE",
    booktitle = "{Workshop on isospin symmetry violation: kaons and beyond -- ISO-BREAK 25}",
    title = "{Nonlinear selection rules due to isospin symmetry}",
    url = "https://indico.cern.ch/event/1557894/contributions/6754662",
    year = "2025",
}

@inproceedings{Kapusta_ISOBREAK25,
    author = "Kapusta, Joseph",
    booktitle = "{Workshop on isospin symmetry violation: kaons and beyond -- ISO-BREAK 25}",
    title = "{A Brief History of Thermal and Statistical Models}",
    url = "https://indico.cern.ch/event/1557894/contributions/6698173",
    year = "2025",
}

@inproceedings{Gorenstein_ISOBREAK25,
    author = "Gorenstein, Mark",
    booktitle = "{Workshop on isospin symmetry violation: kaons and beyond -- ISO-BREAK 25}",
    title = "{Kaon puzzle in statistical model}",
    url = "https://indico.cern.ch/event/1557894/contributions/6698097",
    year = "2025",
}

@inproceedings{Huang_ISOBREAK25,
    author = "Huang, Linqin",
    collaboration = "for BESIII",
    booktitle = "{Workshop on isospin symmetry violation: kaons and beyond -- ISO-BREAK 25}",
    title = "{Charged to Neutral kaon ratio in BESIII}",
    url = "https://indico.cern.ch/event/1557894/contributions/6698145",
    year = "2025",
}

@inproceedings{Lorenz_ISOBREAK25,
    author = "Lorenz, Manuel",
    collaboration = "for HADES",
    booktitle = "{Workshop on isospin symmetry violation: kaons and beyond -- ISO-BREAK 25}",
    title = "{Charged to neutral kaon ratio in HADES}",
    url = "https://indico.cern.ch/event/1557894/contributions/6698091",
    year = "2025",
}

@inproceedings{Dutta_ISOBREAK25,
    author = "Dutta, Dipangkar",
    booktitle = "{Workshop on isospin symmetry violation: kaons and beyond -- ISO-BREAK 25}",
    title = "{Flavor Dependence of Charged Pion Fragmentation Functions}",
    url = "https://indico.cern.ch/event/1557894/contributions/6698146",
    year = "2025",
}

@inproceedings{Stroebele_ISOBREAK25,
    author = "Stroebele, Herbert",
    booktitle = "{Workshop on isospin symmetry violation: kaons and beyond -- ISO-BREAK 25}",
    title = "{Charged to neutral kaon ratio in the NA35 and NA49 experiments}",
    url = "https://indico.cern.ch/event/1557894/contributions/6698086",
    year = "2025",
}

@inproceedings{Kowalski_ISOBREAK25,
    author = "Kowalski, Seweryn and Rybicki, Andrzej",
    booktitle = "{Workshop on isospin symmetry violation: kaons and beyond -- ISO-BREAK 25}",
    title = "{Charged to Neutral Kaon Ratio in NA61/SHINE}",
    url = "https://indico.cern.ch/event/1557894/contributions/6698085",
    year = "2025",
}

@inproceedings{Drachenberg_ISOBREAK25,
    author = "Drachenberg, Jim",
    booktitle = "{Workshop on isospin symmetry violation: kaons and beyond -- ISO-BREAK 25}",
    title = "{Background on Blind Analyses}",
    url = "https://indico.cern.ch/event/1557894/contributions/6698171",
    year = "2025",
}

@article{Margetis:1999ge,
    author = "Margetis, S. and others",
    title = "{Strangeness measurements in NA49 experiment with Pb projectiles}",
    doi = "10.1088/0954-3899/25/2/006",
    journal = "J. Phys. G",
    volume = "25",
    pages = "189--197",
    year = "1999"
}

@inproceedings{Barlow:2002yb,
    author = "Barlow, Roger",
    title = "{Systematic errors: Facts and fictions}",
    booktitle = "{Conference on Advanced Statistical Techniques in Particle Physics}",
    eprint = "hep-ex/0207026",
    archivePrefix = "arXiv",
    reportNumber = "MAN-HEP-02-01",
    pages = "134--144",
    month = "7",
    year = "2002",
}

@article{Forster:2007qk,
    author = "F{\"o}rster, A. and others",
    title = "{Production of $K^+$ and of $K^-$ mesons in heavy-ion collisions from 0.6$A$ to 2.0$A$ GeV incident energy}",
    eprint = "nucl-ex/0701014",
    archivePrefix = "arXiv",
    doi = "10.1103/PhysRevC.75.024906",
    journal = "Phys. Rev. C",
    volume = "75",
    pages = "024906",
    year = "2007"
}

@article{HADES:2023sre,
    author = "Yassine, R. Abou and others",
    collaboration = "HADES",
    title = "{Hadron production and propagation in pion-induced reactions on nuclei}",
    eprint = "2301.03940",
    archivePrefix = "arXiv",
    primaryClass = "nucl-ex",
    doi = "10.1140/epja/s10050-024-01346-y",
    journal = "Eur. Phys. J. A",
    volume = "60",
    number = "7",
    pages = "156",
    year = "2024"
}

@article{HADES:2018qkj,
    author = "Adamczewski-Musch, J. and others",
    collaboration = "HADES",
    title = "{Strong Absorption of Hadrons with Hidden and Open Strangeness in Nuclear Matter}",
    eprint = "1812.03728",
    archivePrefix = "arXiv",
    primaryClass = "nucl-ex",
    doi = "10.1103/PhysRevLett.123.022002",
    journal = "Phys. Rev. Lett.",
    volume = "123",
    number = "2",
    pages = "022002",
    year = "2019"
}

@article{Bratkovskaya:2011wp,
    author = "Bratkovskaya, E. L. and Cassing, W. and Konchakovski, V. P. and Linnyk, O.",
    title = "{Parton--Hadron-String Dynamics at relativistic collider energies}",
    eprint = "1101.5793",
    archivePrefix = "arXiv",
    primaryClass = "nucl-th",
    doi = "10.1016/j.nuclphysa.2011.03.003",
    journal = "Nucl. Phys. A",
    volume = "856",
    pages = "162--182",
    year = "2011"
}

@article{SMASH:2016zqf,
    author = "Weil, J. and others",
    collaboration = "SMASH",
    title = "{Particle production and equilibrium properties within a new hadron transport approach for heavy-ion collisions}",
    eprint = "1606.06642",
    archivePrefix = "arXiv",
    primaryClass = "nucl-th",
    doi = "10.1103/PhysRevC.94.054905",
    journal = "Phys. Rev. C",
    volume = "94",
    number = "5",
    pages = "054905",
    year = "2016"
}

@article{Pierog:2013ria,
    author = "Pierog, T. and Karpenko, Iu. and Katzy, J. M. and Yatsenko, E. and Werner, K.",
    title = "{EPOS LHC: Test of collective hadronization with data measured at the CERN Large Hadron Collider}",
    eprint = "1306.0121",
    archivePrefix = "arXiv",
    primaryClass = "hep-ph",
    reportNumber = "DESY-13-125",
    doi = "10.1103/PhysRevC.92.034906",
    journal = "Phys. Rev. C",
    volume = "92",
    number = "3",
    pages = "034906",
    year = "2015"
}

@article{Price:2023cll,
    author = "Price, Will and Formanek, Martin and Rafelski, Johann",
    title = "{Born-Infeld Nonlinear Electromagnetism in Relativistic Heavy Ion Collisions}",
    eprint = "2306.07704",
    archivePrefix = "arXiv",
    primaryClass = "nucl-th",
    doi = "10.12693/APhysPolA.143.S87",
    journal = "Acta Phys. Polon. A",
    volume = "143",
    pages = "S87",
    year = "2023"
}

@article{Novario:2021low,
    author = "Novario, S. J. and Lonardoni, D. and Gandolfi, S. and Hagen, G.",
    title = "{Trends of Neutron Skins and Radii of Mirror Nuclei from First Principles}",
    eprint = "2111.12775",
    archivePrefix = "arXiv",
    primaryClass = "nucl-th",
    reportNumber = "LA-UR-21-31426",
    doi = "10.1103/PhysRevLett.130.032501",
    journal = "Phys. Rev. Lett.",
    volume = "130",
    number = "3",
    pages = "032501",
    year = "2023"
}

@inproceedings{Giacosa_ISOBREAK25,
    author = "Giacosa, Francesco",
    booktitle = "{Workshop on isospin symmetry violation: kaons and beyond -- ISO-BREAK 25}",
    title = "{Isospin}",
    url = "https://indico.cern.ch/event/1557894/contributions/6698056",
    year = "2025",
}

@inproceedings{Bleicher_ISOBREAK25,
    author = "Bleicher, Marcus",
    booktitle = "{Workshop on isospin symmetry violation: kaons and beyond -- ISO-BREAK 25}",
    title = "{Particle production in string models}",
    url = "https://indico.cern.ch/event/1557894/contributions/6698096",
    year = "2025",
}

@inproceedings{Stepaniak_ISOBREAK25,
    author = "Stepaniak, Johanna",
    booktitle = "{Workshop on isospin symmetry violation: kaons and beyond -- ISO-BREAK 25}",
    title = "{On the hadronisation via quark coalescence}",
    url = "https://indico.cern.ch/event/1557894/contributions/6698098",
    year = "2025",
}

@inproceedings{Vitiuk_ISOBREAK25,
    author = "Vitiuk, Oleksandr",
    booktitle = "{Workshop on isospin symmetry violation: kaons and beyond -- ISO-BREAK 25}",
    title = "{Quantifying Charged Kaon Excess Anomaly
Within Hadronic Transport Approach}",
    url = "https://indico.cern.ch/event/1557894/contributions/6698100",
    year = "2025",
}

@inproceedings{Samanta_ISOBREAK25,
    author = "Samanta, Subhasis",
    booktitle = "{Workshop on isospin symmetry violation: kaons and beyond -- ISO-BREAK 25}",
    title = "{Study of isospin symmetry violation in kaons and beyond using the HRG model}",
    url = "https://indico.cern.ch/event/1557894/contributions/6698101",
    year = "2025",
}

@inproceedings{Rohrmoser_ISOBREAK25,
    author = "Rohrmoser, Martin and Giacosa, Francesco",
    booktitle = "{Workshop on isospin symmetry violation: kaons and beyond -- ISO-BREAK 25}",
    title = "{Coalescence + fit}",
    url = "https://indico.cern.ch/event/1557894/contributions/6698123",
    year = "2025",
}

@inproceedings{Pisarski_ISOBREAK25,
    author = "Pisarski, Robert ",
    booktitle = "{Workshop on isospin symmetry violation: kaons and beyond -- ISO-BREAK 25}",
    title = "{Isospin violation of the axial anomaly}",
    url = "https://indico.cern.ch/event/1557894/contributions/6698170",
    year = "2025",
}

@inproceedings{Rafelski_ISOBREAK25,
    author = "Rafelski, Jan ",
    booktitle = "{Workshop on isospin symmetry violation: kaons and beyond -- ISO-BREAK 25}",
    title = "{On the Possible Origin of Kaon Asymmetry}",
    url = "https://indico.cern.ch/event/1557894/contributions/6744203",
    year = "2025",
}

@inproceedings{Mrowczynski_ISOBREAK25,
    author = "Mr{\'o}wczy{\'n}ski, Stanisław ",
    booktitle = "{Workshop on isospin symmetry violation: kaons and beyond -- ISO-BREAK 25}",
    title = "{Effect of $q$--$\overline{q}$ mutual interaction on $q$--$\overline{q}$ creation}",
    url = "https://indico.cern.ch/event/1557894/contributions/6698190",
    year = "2025",
}

@inproceedings{Pszczel_ISOBREAK25,
    author = "Pszczel, Damian ",
    booktitle = "{Workshop on isospin symmetry violation: kaons and beyond -- ISO-BREAK 25}",
    title = "{Beyond kaons: testing isospin symmetry with pions at NA61/SHINE. Feasibility of $\pi^0$ measurement via photon conversion}",
    url = "https://indico.cern.ch/event/1557894/contributions/6754656",
    year = "2025",
}

@inproceedings{Zhang_ISOBREAK25,
    author = "Zhang, Li'Ang ",
    booktitle = "{Workshop on isospin symmetry violation: kaons and beyond -- ISO-BREAK 25}",
    title = "{Measurement of $K^0_S$ and $K^\pm$ production in Au+Au collisions at the high baryon density region}",
    url = "https://indico.cern.ch/event/1557894/contributions/6698088",
    year = "2025",
}

@inproceedings{Ozvenchuk_ISOBREAK25,
    author = "Ozvenchuk, Vitalii",
    booktitle = "{Private Communication}",
    title = "{Value of $R_K$ in PHSD}",
    year = "2025",
}

@inproceedings{Rohrmoser_private,
    author = "Rohrmoser, Martin",
    booktitle = "{Private Communication}",
    title = "{Value of $R_K$ in QC approach without isospin breaking}",
    year = "2026",
}

@inproceedings{Brandt_ISOBREAK25,
    author = "Bastian B. Brandt ",
    booktitle = "{Workshop on isospin symmetry violation: kaons and beyond -- ISO-BREAK 25}",
    title = "{Strong isospin breaking in lattice QCD}",
    url = "https://indico.cern.ch/event/1557894/contributions/6698095",
    year = "2025",
}

@inproceedings{Luscher:2010ae,
    author = "L{\"u}scher, Martin",
    title = "{Computational Strategies in Lattice QCD}",
    booktitle = "{Les Houches Summer School: Session 93: Modern perspectives in lattice QCD: Quantum field theory and high performance computing}",
    eprint = "1002.4232",
    archivePrefix = "arXiv",
    primaryClass = "hep-lat",
    reportNumber = "CERN-PH-TH-2010-047",
    pages = "331--399",
    month = "2",
    year = "2010"
}

@article{Philipsen:2012nu,
    author = "Philipsen, Owe",
    title = "{The QCD equation of state from the lattice}",
    eprint = "1207.5999",
    archivePrefix = "arXiv",
    primaryClass = "hep-lat",
    doi = "10.1016/j.ppnp.2012.09.003",
    journal = "Prog. Part. Nucl. Phys.",
    volume = "70",
    pages = "55--107",
    year = "2013"
}

@article{FlavourLatticeAveragingGroupFLAG:2024oxs,
    author = "Aoki, Y. and others",
    collaboration = "Flavour Lattice Averaging Group (FLAG)",
    title = "{FLAG review 2024}",
    eprint = "2411.04268",
    archivePrefix = "arXiv",
    primaryClass = "hep-lat",
    reportNumber = "CERN-TH-2024-192, FERMILAB-PUB-24-0785-T",
    doi = "10.1103/nfzp-p5dn",
    journal = "Phys. Rev. D",
    volume = "113",
    number = "1",
    pages = "014508",
    year = "2026"
}

@article{Frezzotti:2022dwn,
    author = "Frezzotti, R. and Gagliardi, G. and Lubicz, V. and Martinelli, G. and Sanfilippo, F. and Simula, S.",
    title = "{Lattice calculation of the pion mass difference $M_{\pi^+}-M_{\pi^0}$ at order $\mathcal{O}(\alpha_\mathrm{em})$}",
    eprint = "2202.11970",
    archivePrefix = "arXiv",
    primaryClass = "hep-lat",
    doi = "10.1103/PhysRevD.106.014502",
    journal = "Phys. Rev. D",
    volume = "106",
    number = "1",
    pages = "014502",
    year = "2022"
}

@article{Gavai:2002fi,
    author = "Gavai, Rajiv V. and Gupta, Sourendu",
    title = "{Phase transition in QCD with broken $SU(2)$ flavor symmetry}",
    eprint = "hep-lat/0208019",
    archivePrefix = "arXiv",
    reportNumber = "TIFR-TH-02-27",
    doi = "10.1103/PhysRevD.66.094510",
    journal = "Phys. Rev. D",
    volume = "66",
    pages = "094510",
    year = "2002"
}

@article{Bierlich:2022pfr,
    author = "Bierlich, Christian and others",
    title = "{A comprehensive guide to the physics and usage of PYTHIA 8.3}",
    eprint = "2203.11601",
    archivePrefix = "arXiv",
    primaryClass = "hep-ph",
    reportNumber = "LU-TP 22-16, MCNET-22-04, FERMILAB-PUB-22-227-SCD",
    doi = "10.21468/SciPostPhysCodeb.8",
    journal = "SciPost Phys. Codeb.",
    volume = "2022",
    pages = "8",
    year = "2022"
}

@article{Sakharov:1948plh,
    author = "Sakharov, Andrei D.",
    title = "{Interaction of the electron and the positron in pair production}",
    reportNumber = "RT-4471",
    doi = "10.1070/PU1991v034n05ABEH002492",
    journal = "Zh. Eksp. Teor. Fiz.",
    volume = "18",
    pages = "631--635",
    year = "1948"
}

@book{Landau:1991wop,
    author = "Landau, Lev Davidovich and Lifshits, E. M.",
    title = "{Quantum Mechanics}: {Non-Relativistic Theory}",
    doi = "10.1016/C2013-0-02793-4",
    isbn = "978-0-7506-3539-4",
    publisher = "Butterworth-Heinemann",
    address = "Oxford",
    series = "Course of Theoretical Physics",
    volume = "v.3",
    year = "1991"
}

@article{Kniehl:2004rk,
    author = "Kniehl, B. A. and Penin, A. A. and Schroder, Y. and Smirnov, Vladimir A. and Steinhauser, M.",
    title = "{Two-loop static QCD potential for general colour state}",
    eprint = "hep-ph/0412083",
    archivePrefix = "arXiv",
    reportNumber = "BI-TP-2004-38, DESY-04-232, SFB-CPP-04-67, TTP-04-24",
    doi = "10.1016/j.physletb.2004.12.024",
    journal = "Phys. Lett. B",
    volume = "607",
    pages = "96--100",
    year = "2005"
}

@article{Gavai:1994in,
    author = "Gavai, R. and Kharzeev, D. and Satz, H. and Schuler, G. A. and Sridhar, K. and Vogt, R.",
    title = "{Quarkonium production in hadronic collisions}",
    eprint = "hep-ph/9502270",
    archivePrefix = "arXiv",
    reportNumber = "CERN-TH-7526-94, BI-TP-63-94, CERN-TH.7526-94",
    doi = "10.1142/S0217751X95001443",
    journal = "Int. J. Mod. Phys. A",
    volume = "10",
    pages = "3043--3070",
    year = "1995"
}

@article{ALICE:2010vtz,
    author = "Aamodt, K. and others",
    collaboration = "ALICE",
    title = "{Strange particle production in proton--proton collisions
at $\sqrt{s}$ = 0.9 TeV with ALICE at the LHC}",
    eprint = "1012.3257",
    archivePrefix = "arXiv",
    primaryClass = "hep-ex",
    reportNumber = "CERN-PH-EP-2010-065",
    doi = "10.1140/epjc/s10052-011-1594-5",
    journal = "Eur. Phys. J. C",
    volume = "71",
    pages = "1594",
    year = "2011"
}

@article{Torrieri:2004zz,
    author = "Torrieri, Giorgio and Steinke, S. and Broniowski, Wojciech and Florkowski, Wojciech and Letessier, Jean and Rafelski, Johann",
    title = "{SHARE: Statistical hadronization with resonances}",
    eprint = "nucl-th/0404083",
    archivePrefix = "arXiv",
    doi = "10.1016/j.cpc.2005.01.004",
    journal = "Comput. Phys. Commun.",
    volume = "167",
    pages = "229--251",
    year = "2005"
}

@article{Torrieri:2006xi,
    author = "Torrieri, G. and Jeon, S. and Letessier, J. and Rafelski, Johann",
    title = "{SHAREv2: Fluctuations and a comprehensive treatment of decay feed-down}",
    eprint = "nucl-th/0603026",
    archivePrefix = "arXiv",
    doi = "10.1016/j.cpc.2006.07.010",
    journal = "Comput. Phys. Commun.",
    volume = "175",
    pages = "635--649",
    year = "2006"
}

@article{Petran:2013dva,
    author = "Petran, M. and Letessier, J. and Rafelski, J. and Torrieri, G.",
    title = "{SHARE with CHARM}",
    eprint = "1310.5108",
    archivePrefix = "arXiv",
    primaryClass = "hep-ph",
    doi = "10.1016/j.cpc.2014.02.026",
    journal = "Comput. Phys. Commun.",
    volume = "185",
    pages = "2056--2079",
    year = "2014"
}

@article{Grayson:2022asf,
     author = "Grayson, Christopher and Formanek, Martin and Rafelski, Johann and Mueller, Berndt",
     title = "{Dynamic magnetic response of the quark-gluon plasma to electromagnetic fields}",
     eprint = "2204.14186",
     archivePrefix = "arXiv",
     primaryClass = "hep-ph",
     doi = "10.1103/PhysRevD.106.014011",
     journal = "Phys. Rev. D",
     volume = "106",
     number = "1",
     pages = "014011",
     year = "2022"
}

@article{Labun:2012jf,
     author = "Labun, Lance and Rafelski, Johann",
     title = "{Acceleration and vacuum temperature}",
     eprint = "1203.6148",
     archivePrefix = "arXiv",
     primaryClass = "hep-ph",
     doi = "10.1103/PhysRevD.86.041701",
     journal = "Phys. Rev. D",
     volume = "86",
     pages = "041701",
     year = "2012"
}

@article{Rafelski:2022bsv,
     author = "Rafelski, Johann and Evans, Stefan and Labun, Lance",
     title = "{Study of QED singular properties for variable gyromagnetic ratio $g\simeq 2$}",
     eprint = "2212.13165",
     archivePrefix = "arXiv",
     primaryClass = "hep-th",
     doi = "10.1103/PhysRevD.107.076002",
     journal = "Phys. Rev. D",
     volume = "107",
     year = "2023"
}

@article{WA98:1998psk,
    author = "Aggarwal, M. M. and others",
    collaboration = "WA98",
    title = "{Centrality Dependence of Neutral Pion Production in 158$A$ GeV $^{208}$Pb + $^{208}$Pb Collisions}",
    eprint = "nucl-ex/9806004",
    archivePrefix = "arXiv",
    reportNumber = "IKP-MS-980601",
    doi = "10.1103/PhysRevLett.81.4087",
    journal = "Phys. Rev. Lett.",
    volume = "81",
    pages = "4087--4091",
    year = "1998",
    note = "[Erratum: Phys.Rev.Lett. 84, 578--579 (2000)]"
}

@article{PHENIX:2012jha,
    author = "Adare, A. and others",
    collaboration = "PHENIX",
    title = "{Neutral pion production with respect to centrality and reaction plane in Au$+$Au collisions at $\sqrt{s_{NN}}$=200 GeV}",
    eprint = "1208.2254",
    archivePrefix = "arXiv",
    primaryClass = "nucl-ex",
    doi = "10.1103/PhysRevC.87.034911",
    journal = "Phys. Rev. C",
    volume = "87",
    number = "3",
    pages = "034911",
    year = "2013"
}

@article{ALICE:2012wos,
    author = "Abelev, B. and others",
    collaboration = "ALICE",
    title = "{Neutral pion and $\eta$ meson production in proton--proton collisions at $\sqrt{s}=0.9$ TeV and $\sqrt{s}=7$ TeV}",
    eprint = "1205.5724",
    archivePrefix = "arXiv",
    primaryClass = "hep-ex",
    reportNumber = "CERN-PH-EP-2012-001",
    doi = "10.1016/j.physletb.2012.09.015",
    journal = "Phys. Lett. B",
    volume = "717",
    pages = "162--172",
    year = "2012"
}

@article{ALICE:2018mdl,
    author = "Acharya, Shreyasi and others",
    collaboration = "ALICE",
    title = "{Neutral pion and $\eta$ meson production at midrapidity in Pb-Pb collisions at $\sqrt{s_{NN}}$ = 2.76 TeV}",
    eprint = "1803.05490",
    archivePrefix = "arXiv",
    primaryClass = "nucl-ex",
    reportNumber = "CERN-EP-2018-040",
    doi = "10.1103/PhysRevC.98.044901",
    journal = "Phys. Rev. C",
    volume = "98",
    number = "4",
    pages = "044901",
    year = "2018"
}

@article{deDivitiis:2011eh,
    author = "de Divitiis, G. M. and others",
    title = "{Isospin breaking effects due to the up-down mass difference in lattice QCD}",
    eprint = "1110.6294",
    archivePrefix = "arXiv",
    primaryClass = "hep-lat",
    reportNumber = "RM3-TH-11-14, RM1-1472-28-10-2011, ROM2F-2011-17, SISSA-58-2011-EP",
    doi = "10.1007/JHEP04(2012)124",
    journal = "JHEP",
    volume = "04",
    pages = "124",
    year = "2012"
}

@article{BMW:2014pzb,
    author = "Borsanyi, Sz. and others",
    collaboration = "BMW",
    title = "{Ab initio calculation of the neutron-proton mass difference}",
    eprint = "1406.4088",
    archivePrefix = "arXiv",
    primaryClass = "hep-lat",
    doi = "10.1126/science.1257050",
    journal = "Science",
    volume = "347",
    pages = "1452--1455",
    year = "2015"
}

@article{Giusti:2017dmp,
    author = "Giusti, D. and Lubicz, V. and Tarantino, C. and Martinelli, G. and Sanfilippo, F. and Simula, S. and Tantalo, N.",
    title = "{Leading isospin-breaking corrections to pion, kaon, and charmed-meson masses with twisted-mass fermions}",
    eprint = "1704.06561",
    archivePrefix = "arXiv",
    primaryClass = "hep-lat",
    reportNumber = "PREPRINT-RM3-TH-17-4",
    doi = "10.1103/PhysRevD.95.114504",
    journal = "Phys. Rev. D",
    volume = "95",
    number = "11",
    pages = "114504",
    year = "2017"
}

@article{West:2020tyo,
    author = "West, Jennifer Rittenhouse",
    title = "{Diquark induced short-range nucleon-nucleon correlations {\&} the EMC effect}",
    eprint = "2009.06968",
    archivePrefix = "arXiv",
    primaryClass = "hep-ph",
    doi = "10.1016/j.nuclphysa.2022.122563",
    journal = "Nucl. Phys. A",
    volume = "1029",
    pages = "122563",
    year = "2023"
}

@article{Jaffe:2004ph,
    author = "Jaffe, R. L.",
    title = "{Exotica}",
    eprint = "hep-ph/0409065",
    archivePrefix = "arXiv",
    reportNumber = "MIT-CTP-3538",
    doi = "10.1016/j.physrep.2004.11.005",
    journal = "Phys. Rept.",
    volume = "409",
    pages = "1--45",
    year = "2005"
}

@inproceedings{Tinti_ISOBREAK25,
    author = "Tinti, Leonardo",
    booktitle = "{Workshop on isospin symmetry violation: kaons and beyond -- ISO-BREAK 25}",
    title = "{Thoughts on isospin}",
    url = "https://indico.cern.ch/event/1557894/contributions/6698192",
    year = "2025",
}

@inproceedings{Ryblewski_ISOBREAK25,
    author = "Ryblewski, Radoslaw",
    booktitle = "{Workshop on isospin symmetry violation: kaons and beyond -- ISO-BREAK 25}",
    title = "{Implementing isospin breaking within THERMINATOR}",
    url = "https://indico.cern.ch/event/1557894/contributions/6698189",
    year = "2025",
}

@inproceedings{Ivanytskyi_ISOBREAK25,
    author = "Ivanytskyi, Oleksii ",
    booktitle = "{Workshop on isospin symmetry violation: kaons and beyond -- ISO-BREAK 25}",
    title = "{Quarkyonic picture of isospin QCD}",
    url = "https://indico.cern.ch/event/1557894/contributions/6698099",
    year = "2025",
}

@article{Ivanytskyi:2025cnn,
    author = "Ivanytskyi, Oleksii",
    title = "{Quarkyonic picture of isospin QCD}",
    eprint = "2505.07076",
    archivePrefix = "arXiv",
    primaryClass = "nucl-th",
    doi = "10.1103/831v-8mp4",
    journal = "Phys. Rev. D",
    volume = "112",
    number = "3",
    pages = "034001",
    year = "2025"
}

@article{ALICE:2019hno,
    author = "Acharya, Shreyasi and others",
    collaboration = "ALICE",
    title = "{Production of charged pions, kaons, and (anti-)protons in Pb-Pb and inelastic $pp$ collisions at $\sqrt{s_{NN}}$ = 5.02 TeV}",
    eprint = "1910.07678",
    archivePrefix = "arXiv",
    primaryClass = "nucl-ex",
    reportNumber = "CERN-EP-2019-208",
    doi = "10.1103/PhysRevC.101.044907",
    journal = "Phys. Rev. C",
    volume = "101",
    number = "4",
    pages = "044907",
    year = "2020"
}

@article{ALICE:2025cqy,
    author = "Abualrob, Ibrahim Jaser and others",
    collaboration = "ALICE",
    title = "{Centrality dependence of strange particle production in Pb--Pb collisions at $\sqrt{s_{\rm NN}} = 5.02$~TeV}",
    eprint = "2511.10360",
    archivePrefix = "arXiv",
    primaryClass = "nucl-ex",
    reportNumber = "CERN-EP-2025-256, CERN-EP-2025-254",
    year = "2025"
}

@article{Meissner:1997ii,
    author = "Mei\ss ner, Ulf G. and Steininger, S.",
    title = "{Isospin violation in pion--nucleon scattering}",
    eprint = "hep-ph/9709453",
    archivePrefix = "arXiv",
    reportNumber = "KFA-IKP-TH-1997-17",
    doi = "10.1016/S0370-2693(97)01418-4",
    journal = "Phys. Lett. B",
    volume = "419",
    pages = "403--411",
    year = "1998"
}

@article{Hoferichter:2009gn,
    author = "Hoferichter, Martin and Kubis, Bastian and Mei\ss ner, Ulf G.",
    title = "{Isospin violation in low-energy pion--nucleon scattering revisited}",
    eprint = "0909.4390",
    archivePrefix = "arXiv",
    primaryClass = "hep-ph",
    reportNumber = "HISKP-TH-09-29, FZJ-IKP-TH-2009-27",
    doi = "10.1016/j.nuclphysa.2009.11.012",
    journal = "Nucl. Phys. A",
    volume = "833",
    pages = "18--103",
    year = "2010"
}

@article{Hoferichter:2009ez,
    author = "Hoferichter, Martin and Kubis, Bastian and Mei\ss ner, Ulf-G.",
    title = "{Isospin breaking in the pion--nucleon scattering lengths}",
    eprint = "0903.3890",
    archivePrefix = "arXiv",
    primaryClass = "hep-ph",
    reportNumber = "HISKP-TH-09-11, FZJ-IKP-TH-2009-9",
    doi = "10.1016/j.physletb.2009.05.068",
    journal = "Phys. Lett. B",
    volume = "678",
    pages = "65--71",
    year = "2009"
}

@article{Colangelo:2025iuq,
    author = "Colangelo, Gilberto and Cottini, Martina and Ruiz de Elvira, Jacobo",
    title = "{Isospin-breaking in the {\ensuremath{\pi}}{\ensuremath{\pi}} scattering amplitude. Part I. Effects due to the pion mass difference}",
    eprint = "2511.08680",
    archivePrefix = "arXiv",
    primaryClass = "hep-ph",
    reportNumber = "IPARCOS-UCM-25-059",
    doi = "10.1007/JHEP02(2026)098",
    journal = "JHEP",
    volume = "02",
    pages = "098",
    year = "2026"
}

@article{Tomiya:2016jwr,
    author = "Tomiya, A. and Cossu, G. and Aoki, S. and Fukaya, H. and Hashimoto, S. and Kaneko, T. and Noaki, J.",
    title = "{Evidence of effective axial $U(1)$ symmetry restoration at high temperature QCD}",
    eprint = "1612.01908",
    archivePrefix = "arXiv",
    primaryClass = "hep-lat",
    doi = "10.1103/PhysRevD.96.034509",
    journal = "Phys. Rev. D",
    volume = "96",
    number = "3",
    pages = "034509",
    year = "2017",
    note = "[Addendum: Phys.Rev.D 96, 079902 (2017)]"
}

@article{Braun:2020mhk,
    author = {Braun, Jens and Leonhardt, Marc and Pawlowski, Jan M. and Rosenbl{\"u}h, Daniel},
    collaboration = "QCD",
    title = "{Chiral and effective $U(1)_{\rm A}$ symmetry restoration in QCD}",
    eprint = "2012.06231",
    archivePrefix = "arXiv",
    primaryClass = "hep-ph",
    year = "2020"
}

@article{Nicola:2020smo,
    author = "Nicola, Angel G{\'o}mez",
    title = "{Light quarks at finite temperature: chiral restoration and the fate of the $U(1)_A$ symmetry}",
    eprint = "2012.13809",
    archivePrefix = "arXiv",
    primaryClass = "hep-ph",
    doi = "10.1140/epjs/s11734-021-00147-4",
    journal = "Eur. Phys. J. ST",
    volume = "230",
    number = "6",
    pages = "1645--1657",
    year = "2021"
}

@article{Aarts:2026kpq,
    author = {Aarts, Gert and Allton, Chris and Bignell, Ryan and J{\"a}ger, Benjamin and Kim, Seyong and Skullerud, Jon-Ivar and Smecca, Antonio},
    title = "{On the effective restoration of $U(1)_A$ symmetry at finite temperature}",
    eprint = "2604.11916",
    archivePrefix = "arXiv",
    primaryClass = "hep-lat",
    year = "2026"
}

@article{OConnell:1995nse,
    author = "O'Connell, Heath Bland and Pearce, B. C. and Thomas, Anthony William and Williams, Anthony Gordon",
    title = "{Rho--Omega Mixing, Vector Meson Dominance and the Pion Form-Factor}",
    eprint = "hep-ph/9501251",
    archivePrefix = "arXiv",
    reportNumber = "ADP-95-1-T-168",
    doi = "10.1016/S0146-6410(97)00044-6",
    journal = "Prog. Part. Nucl. Phys.",
    volume = "39",
    pages = "201--252",
    year = "1997"
}

@article{Giacosa:2024orp,
    author = "Giacosa, Francesco and Kov{\'a}cs, Gy{\H{o}}z{\H{o}} and Kov{\'a}cs, P{\'e}ter and Pisarski, Robert D. and Rennecke, Fabian",
    title = "{Anomalous $U(1)_A$ couplings and the Columbia plot}",
    eprint = "2410.08185",
    archivePrefix = "arXiv",
    primaryClass = "hep-ph",
    doi = "10.1103/PhysRevD.111.016014",
    journal = "Phys. Rev. D",
    volume = "111",
    number = "1",
    pages = "016014",
    year = "2025"
}

@article{Vertesi:2009ca,
    author = "V\'ertesi, R. and Cs{\"o}rg{\H{o}}, T. and Sziklai, J.",
    title = "{Significant in-medium $\eta '$ mass reduction in $\sqrt{s_{NN}}$ $=$ 200 GeV Au+Au collisions}",
    eprint = "0905.2803",
    archivePrefix = "arXiv",
    primaryClass = "nucl-th",
    doi = "10.1016/j.nuclphysa.2009.10.112",
    journal = "Nucl. Phys. A",
    volume = "830",
    pages = "631C--632C",
    year = "2009"
}

@article{Borasoy:2006uv,
    author = "Borasoy, B. and Mei{\ss}ner, Ulf-G. and Ni{\ss}ler, R.",
    title = "{On the extraction of the quark mass ratio $(m_d - m_u)/ m_s$ from $\Gamma(\eta' \rightarrow \pi^0 \pi^+ \pi^-) / \Gamma(\eta ' \rightarrow \eta \pi^+ \pi^-)$}",
    eprint = "hep-ph/0609010",
    archivePrefix = "arXiv",
    reportNumber = "HISKP-TH-06-24, FZJ-IKP-TH-2006-23",
    doi = "10.1016/j.physletb.2006.10.020",
    journal = "Phys. Lett. B",
    volume = "643",
    pages = "41--45",
    year = "2006"
}
